\documentclass[aps,twocolumn,pra,showkeys,superscriptaddress]{revtex4-2}
\usepackage{amsmath, amssymb}
\usepackage{mathrsfs}
\usepackage{array}
\usepackage{booktabs}
\usepackage{tabu}
\usepackage{dcolumn}
\usepackage{amsmath}
\usepackage{amsfonts}
\usepackage{float}
\usepackage{amssymb}
\usepackage{siunitx}
\usepackage{textcomp}
\usepackage{graphicx,color}
\usepackage[colorlinks={true}]{hyperref}
\hypersetup{colorlinks=true,linkcolor=red,citecolor=magenta,urlcolor=blue}
\usepackage{graphicx}
\usepackage{graphicx}
\usepackage{dcolumn}
\usepackage{bm}
\usepackage{pstricks}
\usepackage{braket}
\usepackage{orcidlink}
\usepackage{comment}
\usepackage{enumitem}

\begin{document}

%%%%%%%%%%%%%%%%%%%%%%%%%%%%%%%%%%%%%%

\title{Counterdiabatic Raman Atom Optics for High-Fidelity Sequential Large-Momentum-Transfer Interferometry}

%%%%%%%%%%%%%%%%%%%%%%%%%%%%%%%%%%%%%%

\author{Asad Ali\orcidlink{0000-0001-9243-417X}} \email{asal68826@hbku.edu.qa}
\affiliation{Qatar Center for Quantum Computing, College of Science and Engineering, Hamad Bin Khalifa University, Doha, Qatar}

\author{H.~Arian~Zad\orcidlink{0000-0002-1348-1777}}  
\affiliation{Department of Theoretical Physics and Astrophysics, Faculty of Science of P. J. \v{S}af{\'a}rik University, Park Angelinum 9, 040 01 Ko\v{s}ice, Slovak Republic}

\author{Saif Al-Kuwari\orcidlink{0000-0002-4402-7710}}
\affiliation{Qatar Center for Quantum Computing, College of Science and Engineering, Hamad Bin Khalifa University, Doha, Qatar}

\author{M.~I.~Hussain\orcidlink{0000-0002-6231-7746}}
\affiliation{Qatar Center for Quantum Computing, College of Science and Engineering, Hamad Bin Khalifa University, Doha, Qatar}

\author{M.~T.~Rahim\orcidlink{0000-0003-1529-928X}} 
\affiliation{Qatar Center for Quantum Computing, College of Science and Engineering, Hamad Bin Khalifa University, Doha, Qatar}

\author{H.~Kuniyil\orcidlink{0000-0003-0338-1278}} 
\affiliation{Qatar Center for Quantum Computing, College of Science and Engineering, Hamad Bin Khalifa University,  Doha, Qatar}

\author{Tim Byrnes} 
\affiliation{
	New York University Shanghai; NYU-ECNU Institute of Physics at NYU Shanghai,
	567 West Yangsi Road, Pudong, Shanghai 200126, China
}
\affiliation{
	Center for Quantum and Topological Systems (CQTS),
	NYUAD Research Institute, New York University Abu Dhabi, UAE
}
\affiliation{
	Department of Physics, New York University,
	New York, NY 10003, USA
}

\author{James~Q.~Quach} 
\affiliation{The University of Adelaide, SA 5005, Australia}

\author{Saeed Haddadi\orcidlink{0000-0002-1596-0763}} 
\affiliation{School of Particles and Accelerators, Institute for Research in Fundamental Sciences (IPM), P.O. Box 19395-5531, Tehran, Iran}

%%%%%%%%%%%%%%%%%%%%%%%%%%%%%%%%%%%%%%

\begin{abstract}
Large-momentum-transfer (LMT) atom interferometry provides a route toward enhanced inertial sensitivity in compact quantum sensors, but its scalability is limited by the accumulation of pulse-transfer errors across long Raman pulse sequences. We investigate theoretically the use of stimulated Raman shortcut-to-adiabatic passage (STIRSAP) for high-fidelity sequential LMT atom optics in a Mach-Zehnder interferometer geometry. The counterdiabatic (CD) correction is encoded directly into the Raman pulse envelopes, eliminating the need for auxiliary microwave or radio-frequency control fields. Numerical simulations based on an effective Raman model show that \(1~\mu\mathrm{s}\) STIRSAP pulses achieve single-pulse transfer fidelities of \(F_\pi = 0.99902\) while maintaining negligible pulse-time overhead even at high momentum order. We analyze the resulting tradeoff between interferometric phase enhancement and compound contrast decay and identify an unconstrained shot-noise optimum near \(n\approx270\). Full three-level and open-system simulations confirm the effective description to better than \(10^{-4}\) but show that spontaneous emission contributes \(5.3\times10^{-4}\) per pulse, which lowers the model optimum to \(n\approx170\). A matched-condition benchmark against composite Raman sequences shows that STIRSAP does not achieve the highest peak fidelity among considered sequences but scatters one to two orders of magnitude less at equal pulse duration, which is the relevant figure of merit over the \(4(n-1)\) pulses of an LMT sequence. The analysis further shows that practical operation at extreme LMT order is constrained by wave-packet separation, vibration noise, Doppler detuning, and accumulated systematic effects rather than by pulse duration itself. Within the model studied, these results characterize the scaling of superadiabatic Raman atom optics and identify the physical limitations that govern compact high-order atom interferometers.
\end{abstract}

%%%%%%%%%%%%%%%%%%%%%%%%%%%%%%%%%%%%%%
\maketitle
\section{Introduction}
\label{sec:introduction}

Cold-atom interferometers (CAIs) provide absolute measurements of inertial and gravitational acceleration referenced directly to atomic standards and have reached sub-\(\mu\mathrm{Gal}\) sensitivities (\(1~\mu\mathrm{Gal}=10^{-8}~\mathrm{m/s^2}\)) in laboratory and transportable gravimeters \citep{kasevich1991atomic,peters1999measurement,peters2001high,menoret2018gravity,hu2013demonstration,gillot2016lne,freier2016mobile}. Their long-term stability, intrinsic calibration, and high sensitivity make them attractive for geodesy, underground surveying, inertial navigation, and tests of fundamental physics \citep{bongs2019taking,geiger2020high,tino2021testing}. An important concern, however, is translating the performance achieved in large laboratory systems into compact and field-deployable interferometer architectures. For a Mach-Zehnder atom interferometer, the leading gravitational phase shift scales as
\begin{equation}
	\Delta\Phi = n k_{\mathrm{eff}} g T^2,
\end{equation}
where \(n\) is the large-momentum-transfer (LMT) order, \(k_{\mathrm{eff}}\) is the effective Raman wave vector, and \(T\) is the interrogation time \citep{muller2008atom,cronin2007atom,chiow2009noise}. In compact fountain geometries, the available free-fall distance strongly constrains \(T\), making high-order LMT atom optics the principal route toward enhanced phase sensitivity \citep{mcguirk2000large,chiow2011large,clade2009large,kotru2015large}. A major limitation of sequential LMT interferometry is the accumulation of pulse-transfer errors across long Raman pulse trains. An interferometer of order \(n\) requires an additional \(4(n-1)\) Raman \(\pi\) pulses, so even small single-pulse infidelities compound rapidly and reduce the observable interferometer contrast \citep{berg2015composite,butts2013efficient,saywell2019optimized}. At a single-pulse fidelity near \(99\%\), for example, the expected contrast already falls below \(70\%\) at \(n=10\). Overcoming this fidelity barrier while maintaining short pulse durations is therefore a central challenge for scalable high-order atom interferometry.

Several approaches have been developed to improve Raman transfer robustness, including composite pulses, optimal-control techniques, Bragg diffraction, and adiabatic Raman protocols \citep{berg2015composite,butts2013efficient,saywell2019optimized, bergmann1998coherent,vitanov2017stimulated,weitz1994atomic}. Stimulated Raman adiabatic passage (STIRAP), in particular, provides robustness against pulse area and intensity fluctuations by following an adiabatic dark-state pathway. However, conventional adiabatic protocols generally require pulse durations of order \(10\)-\(100~\mu\mathrm{s}\), which can introduce substantial dead-time overhead when extended to very large LMT order. 
Shortcuts to adiabaticity (STA) provide an alternative route toward robust and ultrafast quantum control \citep{guery2019shortcuts}. In transitionless driving, an additional counterdiabatic (CD) control term suppresses nonadiabatic transitions while preserving the target adiabatic trajectory \citep{berry2009transitionless,demirplak2003adiabatic}. Applied to Raman atom optics, this approach leads to STIRSAP, which has been experimentally demonstrated in \({}^{87}\mathrm{Rb}\) \citep{du2016experimental}. In STIRSAP, the CD correction can be encoded directly into modified Raman pulse envelopes, eliminating the need for auxiliary microwave or radio-frequency control fields.

In this work, we investigate theoretically the application of STIRSAP Raman atom optics to LMT Mach-Zehnder interferometry in a compact atomic fountain. The focus is not on proposing an immediately realizable gravimeter architecture, but rather on analyzing the interplay between superadiabatic pulse fidelity, interferometric scaling, pulse overhead, and practical high-order limitations. The analysis is motivated by the broader question of whether shortcut-to-adiabatic Raman control can provide a scalable path toward ultrafast high-fidelity atom optics for precision quantum sensing.

The main contributions of this work are as follows. In Sec.~\ref{sec:theory},
we develop a closed-form STIRSAP Raman pulse construction in
which the Berry--Demirplak--Rice counterdiabatic correction is incorporated
entirely into the Raman laser envelopes, requiring no auxiliary
microwave or radio-frequency control fields. In Sec.~\ref{sec:single_pulse},
we characterize the resulting single pulse through
transfer-fidelity calculations, robustness maps, Bloch-sphere trajectories, and
beam-splitter calibration, obtaining \(F_\pi=0.99902\) at
\(T_p=1~\mu\mathrm{s}\). In
Secs.~\ref{sec:threelevel}--\ref{sec:benchmark}, we verify the effective model
against full three-level and open-system simulations and benchmark STIRSAP
against optimized composite-pulse Raman sequences (BB1, SCROFULOUS, CORPSE)
under matched conditions, establishing that the shortcut construction scatters
one to two orders of magnitude less light at equal pulse duration the
discriminating figure of merit over an LMT sequence. In Sec.~\ref{sec:lmt}, we assemble these pulses into a sequential LMT Mach--Zehnder model and quantify the trade-off between phase enhancement and compound-fidelity decay, finding an unconstrained shot-noise optimum near \(n\approx270\) and a practical optimum near \(n\approx170\) when spontaneous emission at \(5.3\times10^{-4}\) per pulse is included.
 The same section incorporates
projected technical noise vibration and Raman phase
noise with Allan-deviation estimates and a comparison to representative
experimental gravimeters, showing that technical vibration noise dominates well
before the shot-noise floor is reached at extreme momentum
order. Experimental feasibility is discussed in
Sec.~\ref{sec:feasibility}, the results are placed in the context of existing
LMT approaches in Sec.~\ref{sec:discussion}, and concluding remarks are given in
Sec.~\ref{sec:conclusions}.
\begin{figure}[t]
	\centering
	\includegraphics[width=\columnwidth]{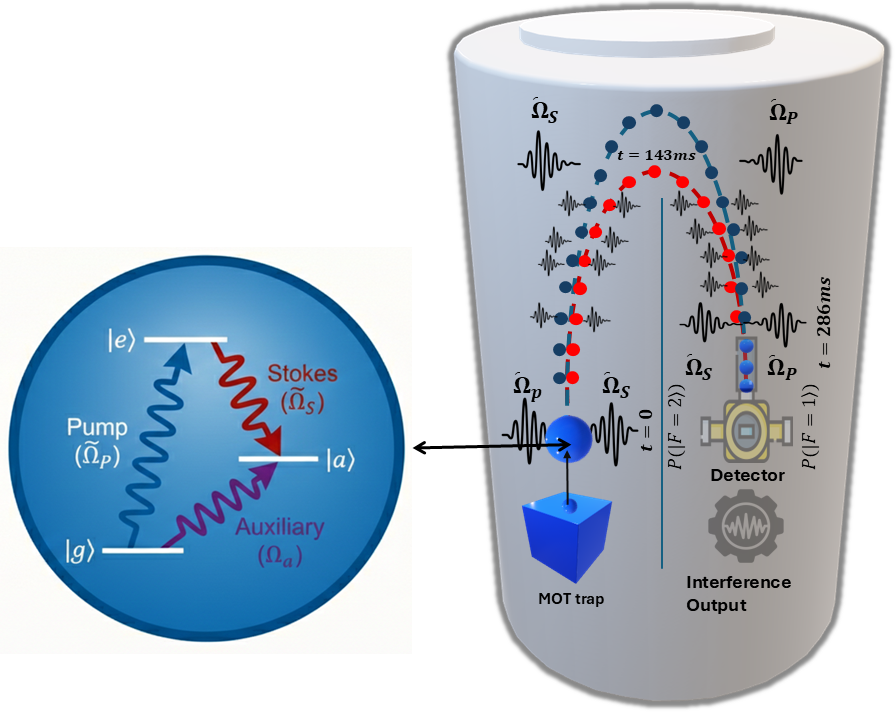}
	\caption{
		Conceptual schematic of the STIRSAP-driven LMT atom interferometer. Three Raman interaction zones at \(t=0\), \(T\), and \(2T\) implement the Mach-Zehnder \(\pi/2\)-\(\pi\)-\(\pi/2\) sequence using counterintuitive STIRSAP pulse pairs coupling \(\ket{g}\) and \(\ket{a}\) through the off-resonant excited state \(\ket{e}\). Sequential Raman \(\pi\) pulses symmetrically accelerate and decelerate the interferometer arms, generating a momentum splitting \(2n\hbar k_{\mathrm{eff}}\). The atomic ensemble is initially prepared in a magneto-optical trap (MOT), which provides a source of ultracold atoms for the interferometric sequence. The auxiliary coupling $\Omega_a$ indicated in the level diagram is shown only to identify the CD term of Eq.~\eqref{eq:Omega_a}; it is not applied as a separate physical field. In the STIRSAP implementation used throughout this work, it is absorbed entirely into the reconstructed optical envelopes $(\widetilde{\Omega}_P,\widetilde{\Omega}_S)$ of Eq.~\eqref{eq:invert}.}
	\label{fig:setup}
\end{figure}

Figure~\ref{fig:setup} illustrates the interferometer concept. Following the initial STIRSAP beam splitter, successive Raman momentum-transfer pulses increase the spatial separation between the interferometer arms to produce a total momentum splitting \(2n\hbar k_{\mathrm{eff}}\). Previous experiments have demonstrated momentum transfer exceeding \(100\hbar k\) using both Raman and optical-clock atom interferometers \citep{chiow2011102,rudolph2020large}, although such systems remain limited by pulse overhead, diffraction efficiency, technical noise, and wave-front control. The present work examines whether shortcut-to-adiabatic Raman shaping can mitigate some of these limitations by combining ultrafast pulse operation with high-fidelity momentum transfer, while simultaneously clarifying the physical constraints that emerge at extreme LMT order.

\section{Theoretical Framework and STIRSAP Raman Control}
\label{sec:theory}

The effective two-level Raman description developed in this section is used to construct the STIRSAP pulses and to model the sequential LMT pulse sequence. Its validity relies on the far-detuned regime
\(\Delta \gg \Omega_{P,S},\) for which transient excited-state population remains perturbatively small. This effective description is subsequently verified against full three-level and open-system simulations in Sec.~\ref{sec:threelevel}, which confirm it to better than \(10^{-4}\) in transfer fidelity at the operating point and quantify the spontaneous-emission channel that the two-level model omits.

\subsection{Effective Raman Hamiltonian}
\label{sec:lambda}

We consider stimulated Raman transitions in \({}^{87}\mathrm{Rb}\) on the D$_2$ line (\(\lambda=780.241~\mathrm{nm}\)) using a three-level \(\Lambda\) configuration consisting of the hyperfine ground states \( \ket{g}\equiv\ket{5S_{1/2},F=1} \) and \( \ket{a}\equiv\ket{5S_{1/2},F=2} \), coupled through the off-resonant excited state \( \ket{e}\equiv\ket{5P_{3/2},F'=2} \). Two counterpropagating Raman fields with angular frequencies \( \omega_P \) and \( \omega_S \) drive the transitions \( \ket{g}\leftrightarrow\ket{e} \) and \( \ket{e}\leftrightarrow\ket{a} \) with single-photon Rabi frequencies \( \Omega_P(t) \) and \( \Omega_S(t) \), respectively. In the rotating-wave approximation and interaction picture, the Hamiltonian is
\begin{equation}
H_\Lambda = \frac{\hbar}{2} \begin{pmatrix} 0 & \Omega_P(t) & 0 \\ \Omega_P(t) & 2\Delta & -\Omega_S(t) \\ 0 & -\Omega_S(t) & 0 \end{pmatrix}, \label{eq:H_lambda}
\end{equation}
where \( \Delta=\omega_{ge}-\omega_P \) is the single-photon detuning. Throughout this work, the Raman transition is assumed to remain on two-photon resonance (\(\delta=0\)) with \( \Delta=2\pi\times5~\mathrm{GHz} \). For peak Raman coupling \( \Omega_0=2\pi\times250~\mathrm{MHz}, \) the detuning ratio \( \Delta/\Omega_0\approx20 \) places the system in the far-detuned Raman regime. Under these conditions, the excited-state population remains perturbatively small and may be adiabatically eliminated to leading order:
\begin{equation}
c_e \approx -\frac{ \Omega_P c_g+\Omega_S c_a }{ 2\Delta }, \label{eq:adiab_elim}
\end{equation}
where, $c_g$, $c_e$, and $c_a$ are the probability amplitudes of the atomic state in the basis $\{|g\rangle\equiv |5S_{1/2}, F=1\rangle, |e\rangle\equiv |5P_{3/2}, F'=2\rangle, |a\rangle\equiv |5S_{1/2}, F=2\rangle\}$ constituting atomic wavefunction as
$|\psi(t)\rangle = c_g(t) |g\rangle + c_e(t) |e\rangle + c_a(t) |a\rangle$.
Substituting Eq.~\eqref{eq:adiab_elim} into the Schr\"odinger equation yields the effective two-level Raman Hamiltonian
\citep{du2017geometric,du2016experimental}
\begin{equation}
H_{\mathrm{eff}}=\frac{\hbar}{2}\begin{pmatrix}-\Delta_{\mathrm{eff}}(t)&\Omega_{\mathrm{eff}}(t)e^{-i\varphi}\\[3pt]\Omega_{\mathrm{eff}}(t)e^{i\varphi}&\Delta_{\mathrm{eff}}(t)\end{pmatrix},\label{eq:Heff}
\end{equation}
where
\begin{equation}
\Delta_{\mathrm{eff}}=\frac{\Omega_P^2-\Omega_S^2}{4\Delta},\qquad\Omega_{\mathrm{eff}}=\frac{\Omega_P\Omega_S}{2\Delta},\label{eq:eff_params}
\end{equation}
and \( \varphi \) is the differential Raman phase. For the parameters used here, \( \Omega_{\mathrm{eff}}^{\max} = 2\pi\times6.25~\mathrm{MHz}, \) corresponding to an effective Raman \(\pi\)-time \(\pi/\Omega_{\mathrm{eff}}^{\max}\) of approximately \(80~\mathrm{ns}\), well below the pulse duration \(T_p=1~\mu\mathrm{s}\) considered throughout this work. The effective two-level treatment provides a computationally efficient description of the sequential LMT pulse sequence. Because the STIRSAP protocol modifies the physical Raman envelopes relative to the original Gaussian profiles, the validity of adiabatic elimination is verified explicitly in Sec.~\ref{sec:threelevel} by full three-level simulations including transient excited-state population and spontaneous scattering; the two descriptions agree to better than \(10^{-4}\) at the operating point.

\subsection{STIRAP Raman Pulses}
\label{sec:stirap}

STIRAP transfers population between the two ground states through a counterintuitive pulse sequence in which the Stokes field precedes the pump field \citep{bergmann1998coherent,vitanov2017stimulated}. Defining the mixing angle
\begin{equation}
\theta(t)=\arctan\!\left[\frac{\Omega_P(t)}{\Omega_S(t)}\right],
\end{equation}
the instantaneous dark state of the three-level Hamiltonian is
\begin{equation}
\ket{D(t)}=\cos\theta(t)\ket{g}-\sin\theta(t)\ket{a}.
\end{equation}

As the pulse ordering evolves from Stokes-dominated to pump-dominated, \( \theta:0\rightarrow\pi/2, \) the dark state adiabatically rotates the population from \( \ket{g} \) to \( \ket{a} \) while ideally suppressing excited-state occupation. Efficient transfer requires the adiabaticity condition \begin{equation} 
	\left| \dot{\theta}(t) \right| \ll \sqrt{ \Omega_P^2(t)+\Omega_S^2(t) }. \label{eq:adiabatic}
\end{equation} 
We employ Gaussian Raman envelopes of width \( \sigma=T_p/6 \) and relative pulse delay \( \tau=0.1\,T_p \): 
\begin{align} 
\Omega_P(t) &= \Omega_0 e^{-(t-t_c-\tau/2)^2/\sigma^2}, \\ \Omega_S(t) &= \Omega_0 e^{-(t-t_c+\tau/2)^2/\sigma^2}, \label{eq:gauss} 
\end{align} 
where \( t_c=T_p/2 \) is the pulse center. For the ultrafast pulse durations considered here, the adiabaticity condition becomes only marginally satisfied. Conventional STIRAP would therefore exhibit nonadiabatic transfer loss, motivating the use of shortcut-to-adiabaticity corrections.
\subsection{Counterdiabatic Raman Control}
\label{sec:sta}
Shortcuts to adiabaticity suppress nonadiabatic transitions by augmenting the adiabatic dynamics with a CD correction \citep{berry2009transitionless,demirplak2003adiabatic,guery2019shortcuts}; recent studies have further examined STA robustness and control-error sensitivity in two- and three-level systems \citep{xu2023quantum} and hybrid spin-superconducting systems~\citep{PhysRevA.111.012621}. For a Hamiltonian with instantaneous eigenstates
\( \{\ket{n(t)}\}, \) the general CD Hamiltonian is \begin{equation} 
H_{\mathrm{CD}} = i\hbar \sum_n \left( \ket{\partial_t n}\bra{n} - \braket{n|\partial_t n} \ket{n}\bra{n} \right). \label{eq:H_CD} 
\end{equation} 
Applied to the effective Raman Hamiltonian~\eqref{eq:Heff}, the CD correction introduces an auxiliary coupling \begin{equation} 
\Omega_a(t) = \frac{ 2\left( \dot{\Omega}_P\Omega_S - \Omega_P\dot{\Omega}_S \right) }{ \Omega_P^2+\Omega_S^2 }. \label{eq:Omega_a} 
\end{equation} 
The resulting superadiabatic Hamiltonian becomes \begin{equation} 
H_{\mathrm{tot}} = \frac{\hbar}{2} \begin{pmatrix} -\Delta_{\mathrm{eff}} & \Omega_{\mathrm{eff}}-i\Omega_a \\ \Omega_{\mathrm{eff}}+i\Omega_a & \Delta_{\mathrm{eff}} \end{pmatrix}, \label{eq:H_total} 
\end{equation} 
where the CD term appears as an additional quadrature coupling with relative phase \(\pi/2\).

In the idealized effective model, the CD correction suppresses
nonadiabatic transfer errors and enables rapid Raman population transfer while
preserving the adiabatic eigenstate trajectory. In practice, however, the
achievable fidelity remains constrained by spontaneous emission, finite laser
power, pulse distortions, Doppler broadening, and the validity of adiabatic
elimination in the ultrafast regime.

\subsection{Superadiabatic Raman Pulse Reconstruction}
\label{sec:modified_params}

To implement the superadiabatic Hamiltonian~\eqref{eq:H_total}
using experimentally accessible Raman fields, the auxiliary CD
coupling is absorbed into modified effective Raman parameters through the
rotation angle
\begin{equation}
	\phi(t)
	=
	\arctan\!\left[
	\frac{\Omega_a(t)}
	{\Omega_{\mathrm{eff}}(t)}
	\right].
\end{equation}

The corresponding superadiabatic coupling and detuning are
\begin{align}
	\widetilde{\Omega}(t)
	&=
	\sqrt{
		\Omega_{\mathrm{eff}}^2(t)+\Omega_a^2(t)
	},
	\label{eq:Otilde}
	\\
	\widetilde{\Delta}(t)
	&=
	\Delta_{\mathrm{eff}}(t)+\dot{\phi}(t).
	\label{eq:Dtilde}
\end{align}

These effective parameters are then inverted algebraically to reconstruct the
physical Raman envelopes
\citep{du2016experimental}
\begin{align}
	\widetilde{\Omega}_P
	&=
	\sqrt{
		2\Delta
		\left(
		\sqrt{
			\widetilde{\Delta}^2+\widetilde{\Omega}^2
		}
		+
		\widetilde{\Delta}
		\right)
	},
	\\
	\widetilde{\Omega}_S
	&=
	\sqrt{
		2\Delta
		\left(
		\sqrt{
			\widetilde{\Delta}^2+\widetilde{\Omega}^2
		}
		-
		\widetilde{\Delta}
		\right)
	}.
	\label{eq:invert}
\end{align}

The reconstructed pulse pair
\(
(\widetilde{\Omega}_P,\widetilde{\Omega}_S)
\)
constitutes the STIRSAP Raman sequence used throughout this work. Importantly,
the CD correction is implemented entirely through optical shaping
of the Raman pulse envelopes and does not require auxiliary microwave or
radio-frequency couplings. Because the reconstructed Raman envelopes differ from the original
Gaussian STIRAP pulses, the modified fields enhance the instantaneous Raman
coupling near the pulse overlap region. The validity of the effective Raman
approximation is verified explicitly against full three-level simulations in
Sec.~\ref{sec:threelevel}, where the largest reconstructed Rabi frequency is
found to remain below $\Delta/14$ throughout the parameter range studied.

The pulse reconstruction is verified by recomputing the effective Raman
parameters generated by the modified physical envelopes:
\begin{equation}
\Delta_{\mathrm{eff}}^{\mathrm{chk}}=\frac{\widetilde{\Omega}_P^2-\widetilde{\Omega}_S^2}{4\Delta},\qquad\Omega_{\mathrm{eff}}^{\mathrm{chk}}=\frac{\widetilde{\Omega}_P\widetilde{\Omega}_S}{2\Delta}.
\end{equation}
The reconstructed parameters reproduce
\((\widetilde{\Delta},\widetilde{\Omega})\)
to numerical precision, confirming the internal consistency of the inversion
within the effective Raman model.

\section{Single-Pulse Performance}
\label{sec:single_pulse}

\subsection{Transfer Fidelity and Pulse Optimization}
\label{sec:fidelity}

The STIRSAP Raman dynamics are propagated numerically using a
piecewise-constant matrix-exponential scheme with \(4\times10^3\) time steps
per pulse duration \(T_p\), which is exactly unitary and treats the fast
rotation at the effective detuning accurately.
At each time step the reconstructed STIRSAP Raman envelopes
\(
(\widetilde{\Omega}_P,\widetilde{\Omega}_S)
\)
are computed from
Eqs.~\eqref{eq:Otilde}-\eqref{eq:invert}
and inserted into the effective Hamiltonian~\eqref{eq:H_total}.

\begin{figure*}[htbp]
	\centering
	\includegraphics[width=0.8\textwidth]{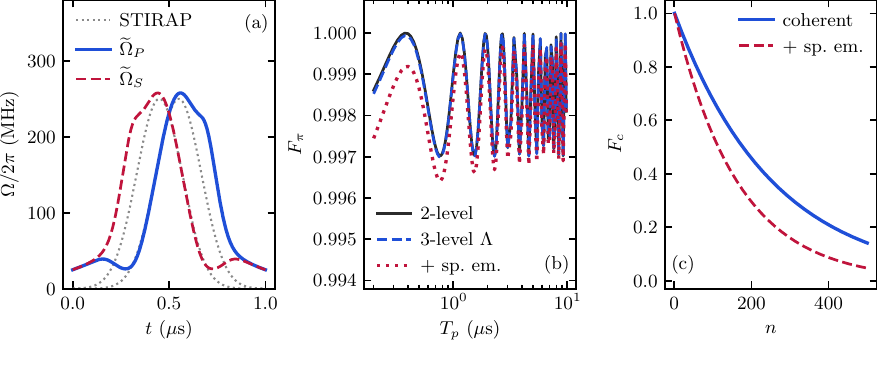}
    \vspace{-0.75cm}
	\caption{
		Single-pulse STIRSAP performance.
		\textbf{(a)}
		Comparison between the original Gaussian STIRAP Raman envelopes (dashed)
		and the reconstructed STIRSAP Raman envelopes (solid) for
		\(T_p=1~\mu\mathrm{s}\).
		\textbf{(b)}
		Raman
		\(\pi\)-pulse fidelity
		\(F_\pi\)
		versus pulse duration.
		High-fidelity transfer persists throughout the ultrafast regime, although
		the oscillatory structure indicates sensitivity to the detailed pulse
		dynamics.
		\textbf{(c)}
		Compound interferometer fidelity
		\(F_c=F_\pi^{4(n-1)}\)
		versus LMT order
		\(n\).
		The exponential accumulation of pulse-transfer errors progressively reduces
		the achievable interferometer contrast at high momentum order in both coherent (solid blue) and with spontaneous emission (dashed red).}
	\label{fig:protocol}
\end{figure*}

Figure~\ref{fig:protocol}(a) compares the original Gaussian STIRAP envelopes
with the reconstructed STIRSAP Raman pulses for
\(
T_p=1~\mu\mathrm{s}.
\)
The superadiabatic correction substantially reshapes the pulse sequence near the
pulse overlap region, where the CD contribution is largest.

The achieved fidelity indicates that superadiabatic Raman shaping can suppress
nonadiabatic transfer errors even in the ultrafast pulse regime. A
matched-condition benchmark against standard composite Raman sequences is
presented in Sec.~\ref{sec:benchmark}.

Figure~\ref{fig:protocol}(b) shows the dependence of the transfer fidelity on pulse duration over the range \( T_p\in[0.2,10]~\mu\mathrm{s}. \) The simulations predict \( F_\pi>99.7\% \) throughout this interval in the coherent model (the minimum falls to \(99.64\%\) once spontaneous emission (sp.~em.) is included), although the fidelity exhibits oscillatory behavior rather than monotonic degradation with decreasing pulse duration. This behavior reflects the interplay between nonadiabatic dynamics and the CD correction in the ultrafast regime. At short pulse durations, the STIRSAP reconstruction modifies the physical Raman envelopes relative to the original Gaussian STIRAP pulses, particularly near the pulse overlap region where the auxiliary coupling \( \Omega_a(t) \) becomes largest. 

We have verified this regime directly by full three-level
and open-system simulation (Sec.~\ref{sec:threelevel}): the reconstructed
Rabi frequencies remain below $\Delta/14$ over the entire range
$T_p\in[0.2,10]~\mu\mathrm{s}$, the transient excited-state population stays
below $5\times10^{-4}$, and the three-level transfer fidelity agrees with the
effective two-level result to better than $10^{-4}$. The oscillatory
dependence of $F_\pi$ on $T_p$ visible in Fig.~\ref{fig:protocol}(b) is a
boundary-truncation artefact rather than a breakdown of adiabatic
elimination, and its amplitude is set by a single geometric quantity that can
be derived in closed form.
The Gaussian envelopes
$\Omega_{P,S}(t)=\Omega_0\exp[-(t-t_{P,S})^2/\sigma^2]$ with $\sigma=T_p/6$
are truncated at the finite window $[0,T_p]$, so the mixing angle
$\theta(t)=\arctan[\Omega_P(t)/\Omega_S(t)]$ does not reach its ideal
boundary values of $0$ and $\pi/2$.
Evaluating the tails at $t=0$ gives $\Omega_P(0)/\Omega_0=1.86\times10^{-5}$
and $\Omega_S(0)/\Omega_0=6.82\times10^{-4}$, yielding the residual mixing
angle
\begin{align}
\theta(0)&=\arctan\!\left[\frac{\Omega_P(0)}{\Omega_S(0)}\right]
=\arctan\!\left[\exp\!\left(-\frac{t_P^2-t_S^2}{\sigma^2}\right)\right]
\nonumber\\
&\simeq 0.0273~\mathrm{rad};
\label{eq:theta0_main}
\end{align}
crucially, $\theta(0)$ depends only on the fixed ratios
$\sigma/T_p=1/6$ and $\delta f=1/10$, so it is the same
for every $T_p$ in the scan.
Time-reversal symmetry of the delay geometry gives
$\theta(T_p)=\pi/2-\theta(0)$ exactly.
Because $\theta(0)\neq0$, the initial state $|g\rangle$ is not exactly the
dark state $|D(0)\rangle=\cos\theta(0)|g\rangle-\sin\theta(0)|a\rangle$ at
the pulse opening; decomposing in the instantaneous dark-bright basis,
$|g\rangle=\cos\theta(0)|D(0)\rangle+\sin\theta(0)|B(0)\rangle$,
where $|B(0)\rangle=\sin\theta(0)|g\rangle+\cos\theta(0)|a\rangle$,
the dark component $\cos\theta(0)|D(0)\rangle$ is carried exactly to
$-\cos\theta(0)|a\rangle$ by the STIRSAP protocol, while the bright component
$\sin\theta(0)|B(0)\rangle$ undergoes oscillatory evolution and accumulates a
$T_p$-dependent phase
$\Phi(T_p)\sim\int_0^{T_p}\!\Omega_{\mathrm{eff}}(t)\,dt\propto T_p$.
Using $|B(T_p)\rangle=\cos\theta(0)|g\rangle+\sin\theta(0)|a\rangle$ from
the boundary symmetry, the bright component contributes
$\sin^2\theta(0)\,e^{i\Phi}$ to $c_a$, giving the total excited amplitude
$c_a=-\cos\theta(0)+\sin^2\!\theta(0)\,e^{i\Phi}$ and therefore
\begin{equation}
F_\pi(T_p)=|c_a|^2
\approx 1-\sin^2\!\theta(0)-2\sin^2\!\theta(0)\cos\Phi(T_p),
\label{eq:Fpi_osc_main}
\end{equation}
where the term $\sin^4\theta(0)$ has been dropped for $\theta(0)\ll1$.
Since $\Omega_{\mathrm{eff}}\propto\Omega_0^2/\Delta$ is independent of $T_p$
while the accumulated phase grows as $\Phi\propto T_p$, the factor $\cos\Phi$
sweeps between $-1$ and $+1$ as $T_p$ varies, producing a peak-to-peak
oscillation
\begin{align}
\Delta F_\pi&=4\sin^2\!\theta(0)
=4\sin^2\!\!\left[\arctan\frac{\Omega_P(0)}{\Omega_S(0)}\right]
\nonumber\\
&\simeq 2.98\times10^{-3},
\label{eq:DeltaFpi_main}
\end{align}
in quantitative agreement with the numerical scan (observed peak-to-peak
$2.982\times10^{-3}$, ratio $1.00$ to three significant figures).
Since the dark-bright decomposition is a property of $\theta(0)$ alone and
is independent of whether the excited state is retained or adiabatically
eliminated, this oscillation is reproduced identically by both the two-level
and the full three-level model.

Figure~\ref{fig:protocol}(c) shows the compound fidelity
$F_c(n)=F_\pi^{\,4(n-1)}$ accumulated over the $4(n-1)$ augmentation pulses
of an order-$n$ sequence, for the coherent value $F_\pi=0.99902$ (solid) and
for $F_\pi=0.99847$ with scattered atoms treated as lost (dashed). Although the infidelity of a single pulse is only \(9.8\times10^{-4}\), the cumulative effect of applying a sequence of \(\sim 10^3\) pulses leads to a substantial reduction in \(F_c\), reaching values of approximately \(0.35\) at \(n\approx270\) and \(0.14\) at \(n=500\). Because the fringe contrast scales directly with \(F_c\) [Eq.~\eqref{eq:contrast}], this decay counteracts the linear enhancement of the phase accumulation and ultimately determines the optimal sensitivity, which will be analyzed in Sec.~\ref{sec:lmt}.

\subsection{Pulse Construction, Robustness, and Single-Pulse Characterization}
\label{sec:robustness}

\begin{figure*}[tp]
	\centering
	\includegraphics[width=0.75\textwidth]{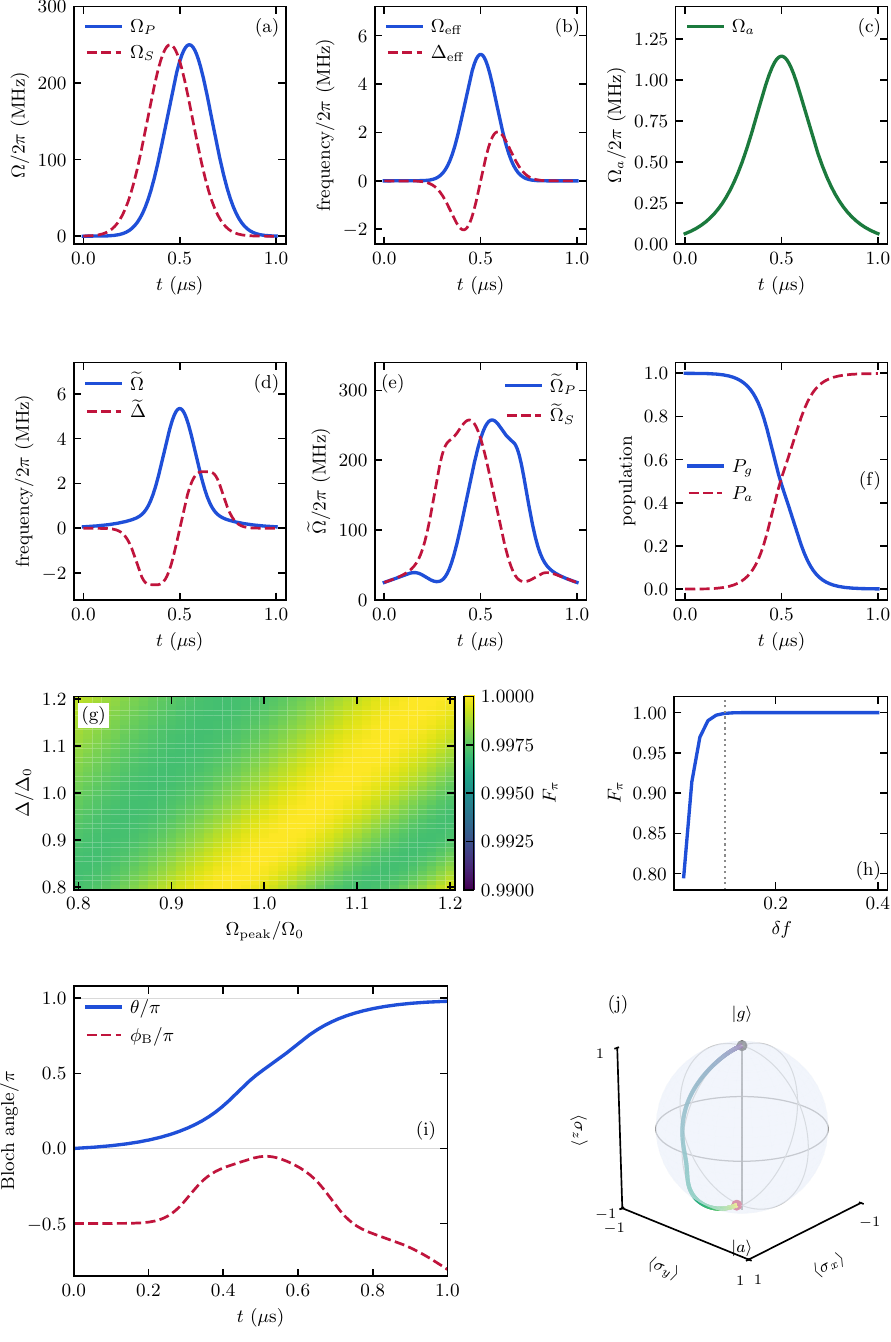}
    \vspace{-0.75cm}
	\caption{
		STIRSAP pulse construction and single-pulse characterization.
		\textbf{(a)}
		Base Gaussian STIRAP Raman envelopes.
		\textbf{(b)}
		Effective Raman coupling and detuning obtained after adiabatic elimination.
		\textbf{(c)}
		CD auxiliary coupling
		\(\Omega_a(t)\).
		\textbf{(d)}
		Modified superadiabatic effective parameters
		\((\widetilde{\Delta},\widetilde{\Omega})\).
		\textbf{(e)}
		Final reconstructed STIRSAP Raman envelopes.
		\textbf{(f)}
		Population dynamics during a full Raman
		\(\pi\)
		pulse, yielding
		\(F_\pi=0.99902\).
		\textbf{(g)}
		Robustness map of the transfer fidelity
		\(F_\pi\)
		under simultaneous variations of the peak Raman coupling and
		single-photon detuning. High fidelity is maintained primarily along
		correlated variations preserving the ratio
		\(\Omega_{\mathrm{peak}}^2/\Delta\).
		\textbf{(h)}
		Transfer fidelity versus relative pulse delay. The operating point used
		throughout this work, \(\delta f=0.1\) (dotted vertical line), lies within a
		broad high-fidelity plateau.
		\textbf{(i)}
		Evolution of the Bloch-sphere polar angle ($\theta$, solid line) and the azimuthal
		angle ($\phi_{\mathrm{B}}$, dashed line) in the superadiabatic frame during the pulse; the
		polar angle $\theta$ sweeps smoothly from \(0\) to \(\pi\) while azimuthal
		angle $\phi_{\mathrm{B}}$
		remains bounded, the signature of adiabatic following.
		\textbf{(j)}
		Bloch-sphere trajectory connecting \(\ket{g}\) (north pole) and
		\(\ket{a}\) (south pole).}
	\label{fig:construction}
\end{figure*}

Figure~\ref{fig:construction} summarizes the complete STIRSAP pulse
construction and its single-pulse characterization. 
Figure~\ref{fig:construction}(a) shows the base counterintuitive STIRAP Raman
envelopes, where the Stokes pulse precedes the pump pulse to establish the
adiabatic transfer pathway. Adiabatic elimination of the off-resonant excited
state produces the effective two-level Raman coupling and detuning shown in
Fig.~\ref{fig:construction}(b). The corresponding CD correction required for
transitionless evolution is displayed in Fig.~\ref{fig:construction}(c), where
the auxiliary coupling \(\Omega_a(t)\) is localized near the temporal overlap
region of the Raman pulses, precisely where nonadiabatic transitions are
strongest. Incorporating this correction modifies both the effective Raman
coupling and detuning [Fig.~\ref{fig:construction}(d)], thereby reshaping the
adiabatic pathway while preserving the desired population transfer. Finally,
inversion of the effective Raman relations yields the physical STIRSAP Raman
envelopes shown in Fig.~\ref{fig:construction}(e). Relative to the original
STIRAP pulses, the reconstructed envelopes exhibit asymmetric temporal
distortions and enhanced overlap, encoding the CD correction entirely within the
optical control fields. This reconstruction enables superadiabatic population
transfer without introducing auxiliary microwave or radio-frequency couplings.

The resulting population dynamics [Fig.~\ref{fig:construction}(f)] show
population transfer from \(\ket{g}\) to \(\ket{a}\) with a
residual initial-state population of \(9.8\times10^{-4}\). The transfer
fidelity is
\begin{equation}
F_\pi = 1-\epsilon_\pi = 0.99902, \qquad \epsilon_\pi = 9.8\times10^{-4}.
\end{equation}
A Raman \(\pi/2\) beam splitter is generated by truncating this
\(\pi\)-pulse sequence at the time for which the target-state population reaches
\(P_a=0.5\); the optimal truncation time is \(t_{\mathrm{bs}}=0.494~\mu\mathrm{s}\),
yielding \(P_a=0.4989\), a beam-splitter imbalance below \(0.23\%\).

The robustness of the STIRSAP transfer to simultaneous variations in Raman
intensity and single-photon detuning is shown in
Fig.~\ref{fig:construction}(g). Fidelity remains above \(0.999\) across a broad
diagonal region in parameter space corresponding approximately to constant
effective Raman coupling \(\Omega_{\mathrm{eff}}\propto\Omega_{\mathrm{peak}}^2/\Delta\),
which follows directly from Eq.~\eqref{eq:eff_params} evaluated
near the pulse overlap, where \(\Omega_P\approx\Omega_S\approx\Omega_{\mathrm{peak}}\).
The results therefore indicate robustness primarily against correlated
variations of Raman intensity and detuning rather than against independent
fluctuations of either parameter alone. At the boundaries of the parameter
space, independent drift in laser power or detuning can reduce the transfer
fidelity substantially, so that maintaining stable high-order LMT operation
would require active stabilization of both Raman intensity and Raman detuning
across the full pulse sequence.

Figure~\ref{fig:construction}(h) shows the dependence of the transfer fidelity
on the STIRAP delay fraction \(\delta f\). The fidelity exhibits a broad
high-fidelity plateau away from the fully overlapping-pulse limit
(\(\delta f\rightarrow0\)), where nonadiabatic coupling becomes strongest. The
operational point adopted throughout this work, \(\delta f=0.1\), lies within
this plateau and yields \(F_\pi\approx0.999\), while the unconstrained optimum
occurs near \(\delta f\approx0.35\). The operating value
\(\delta f=0.1\) is retained rather than \(\delta f\approx0.35\) because the
larger delay increases the counterdiabatic amplitude \(\Omega_a\) and hence the
per-pulse spontaneous-emission loss quantified in Sec.~\ref{sec:threelevel},
which is the figure of merit that governs the sequential LMT interferometer.

Additional insight is provided by the Bloch-sphere representation.
Figure~\ref{fig:construction}(i) shows the evolution of the polar angle $\theta$ and the azimuthal angle $\phi_{\mathrm{B}}$, which vary smoothly throughout
the pulse without abrupt phase excursions. The corresponding trajectory on the
Bloch sphere [Fig.~\ref{fig:construction}(j)] connects the north-pole state
\(\ket{g}\) to the south-pole state \(\ket{a}\) along a continuous path that
closely follows the instantaneous adiabatic eigenstate. The
absence of large oscillations or loops is consistent with suppression of
nonadiabatic transitions, indicating that the STIRSAP reconstruction reproduces
adiabatic-like evolution on a sub-microsecond timescale within this model.

The present parameter scan should be interpreted as a
first-order robustness assessment within the effective Raman model. Additional
effects relevant to high-order interferometry, including Doppler detuning,
spatial pulse-area variation, wave-front inhomogeneity, diffraction phases, and
coherent parasitic diffraction channels, are not included and may substantially
alter the effective robustness under realistic experimental conditions.

\subsection{Three-level and open-system verification}
\label{sec:threelevel}

To test the effective Raman description, we propagate the full three-level
Hamiltonian~\eqref{eq:H_lambda} directly, driven by the reconstructed
envelopes $(\widetilde{\Omega}_P,\widetilde{\Omega}_S)$ at exact two-photon
resonance and without adiabatic elimination. The differential light shift
that plays the role of $\widetilde{\Delta}$ is generated by the envelopes
themselves, so no additional modulation is required. Propagation uses a
piecewise-constant matrix-exponential scheme with $4\times10^3$ steps per
pulse, which treats the fast rotation at $\Delta$ exactly.

\begin{figure*}[t]
	\centering
	\includegraphics[width=0.8\textwidth]{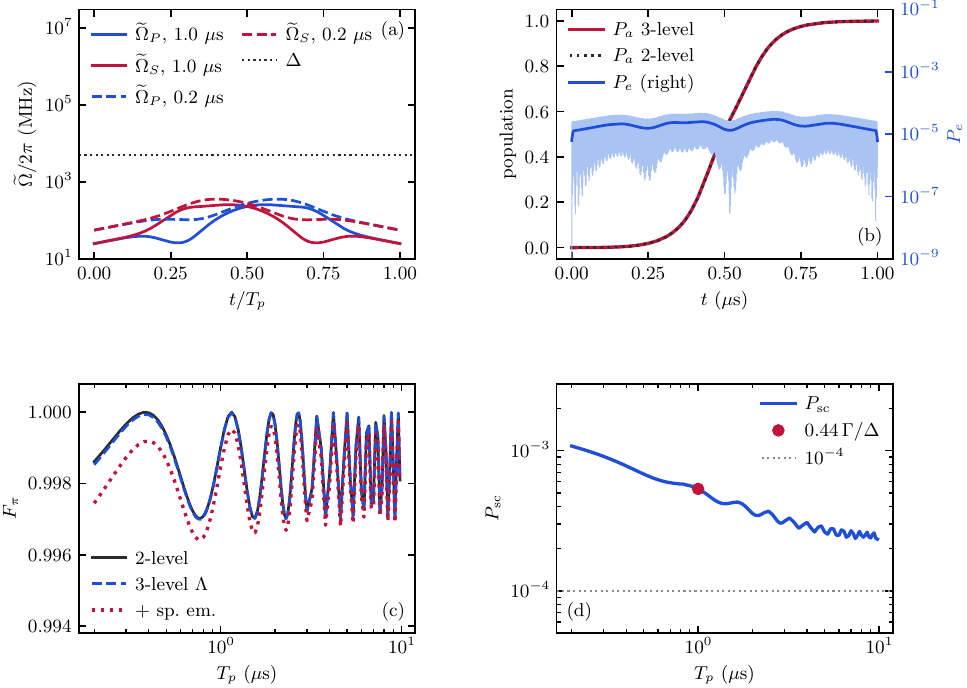}
	\caption{
		Three-level and open-system verification of the STIRSAP pulses.
		\textbf{(a)} Reconstructed Raman envelopes for $T_p=1~\mu\mathrm{s}$
		(solid) and $0.2~\mu\mathrm{s}$ (dashed) against the single-photon
		detuning $\Delta$ (dotted); the envelopes remain below $\Delta/14$.
		\textbf{(b)} Populations during a $\pi$ pulse at $T_p=1~\mu\mathrm{s}$:
		target-state population from the three-level (red) and two-level
		(black dotted) models, with the transient excited-state population
		$P_e$ on the right axis (blue; light trace raw, dark trace averaged
		over $2\pi/\Delta$).
		\textbf{(c)} Transfer fidelity versus pulse duration for the effective
		two-level model, the unitary three-level model, and the three-level
		model with spontaneous emission.
		\textbf{(d)} Spontaneous-emission probability per pulse versus $T_p$,
		with the $0.44\,\Gamma/\Delta$ value at $T_p=1~\mu\mathrm{s}$ and the
		original $10^{-4}$ estimate for reference.}
	\label{fig:threelevel}
\end{figure*}

Figure~\ref{fig:threelevel}(a) shows the reconstructed envelopes for
$T_p=1$ and $0.2~\mu\mathrm{s}$ together with $\Delta$. The largest
instantaneous Rabi frequency is $2\pi\times258~\mathrm{MHz}$ at
$T_p=1~\mu\mathrm{s}$ and $2\pi\times358~\mathrm{MHz}$ at
$T_p=0.2~\mu\mathrm{s}$, so that
$\Delta/\max(\widetilde{\Omega}_P,\widetilde{\Omega}_S)\geq14$ throughout;
the envelopes do not approach the single-photon detuning. The corresponding
populations are shown in Fig.~\ref{fig:threelevel}(b): the transient
excited-state population remains below $5.4\times10^{-5}$ at
$T_p=1~\mu\mathrm{s}$, and the three-level transfer fidelity
$F_\pi^{(3)}=0.99899$ agrees with the effective-model value $F_\pi=0.99902$
to within $2.6\times10^{-5}$. The two models agree to better than $10^{-4}$
for all $T_p\in[0.2,10]~\mu\mathrm{s}$ [Fig.~\ref{fig:threelevel}(c)],
confirming that adiabatic elimination is adequate for the coherent STIRSAP
dynamics in the ultrafast regime.

It is also useful to test the premise motivating the CD
correction. Propagating the uncorrected Gaussian STIRAP envelopes of
Eq.~\eqref{eq:gauss} through the same three-level open-system model, at the
same durations and peak coupling, gives transfer fidelities of $0.733$ at
$T_p=1~\mu\mathrm{s}$ and $0.877$ at $T_p=10~\mu\mathrm{s}$, against $0.998$
and $0.997$ for STIRSAP. The uncorrected values are not monotonic in $T_p$:
for these envelopes the effective two-photon pulse area grows linearly with
duration ($2.2\pi$ at $1~\mu\mathrm{s}$, $21.8\pi$ at $10~\mu\mathrm{s}$), so
plain STIRAP oscillates rather than converging to unit transfer at the
parameters used here. The improvement reported above is therefore attributable
to the CD correction rather than to the choice of envelope.

The effect of spontaneous emission on shortcut-to-adiabaticity protocols in three-level systems has been studied previously \citep{shi2021effect}; here we quantify it directly for the reconstructed STIRSAP envelopes. Spontaneous emission from $\ket{e}$ is included both as a non-Hermitian
decay at the natural rate $\Gamma=2\pi\times6.07~\mathrm{MHz}$, with
scattered atoms removed from the interferometer, and through a Lindblad
master equation with decay back into $\ket{g}$ and $\ket{a}$. The scattering
probability per STIRSAP $\pi$ pulse is
\begin{equation}
P_{\mathrm{sc}}\simeq0.44\,\frac{\Gamma}{\Delta}=5.3\times10^{-4}
\label{eq:Psc}
\end{equation}
at $\Delta=2\pi\times5~\mathrm{GHz}$, found numerically to be exactly
proportional to $1/\Delta$ at fixed $\Omega_{\mathrm{eff}}$ and only weakly
dependent on $T_p$ [Fig.~\ref{fig:threelevel}(d)]. This value is about five
times larger than the estimate $\sim10^{-4}$ used in our initial feasibility
discussion. It arises because the CD correction holds the atom
on the dark state of the unmodified Raman fields, which is a bright-state
admixture of the reconstructed fields of weight $\sim\Omega_a/\widetilde{\Omega}$,
giving an excited-state population of order $\Omega_a/2\Delta$ where the
CD term dominates. A complementary analysis casts the
spontaneous-scattering loss entirely in terms of this bright-state amplitude
and shows that the CD point cancels the bright-state source
exactly, for arbitrary detuning, Rabi frequency, and pulse duration
\citep{ali2026brightstate}; the residual per-pulse scattering quoted here is
the finite-$T_p$, finite-detuning remainder of that cancellation. For comparison, a resonant Raman $\pi$ pulse
with $\Omega_P=\Omega_S=\Omega_0$ at the same detuning scatters with
probability $\pi\Gamma/2\Delta=1.9\times10^{-3}$, roughly four times more,
because it does not follow a dark state. Treating scattered atoms as lost,
the single-pulse fidelity becomes $F_\pi=0.99847$; the Lindblad treatment
gives $0.99874$. This lower scattering at matched detuning and pulse
duration, rather than a higher coherent fidelity, is the specific advantage
offered by the STIRSAP construction over a bare resonant Raman pulse of equal
area.

\begin{figure*}[t]
	\centering
	\includegraphics[width=0.9\textwidth]{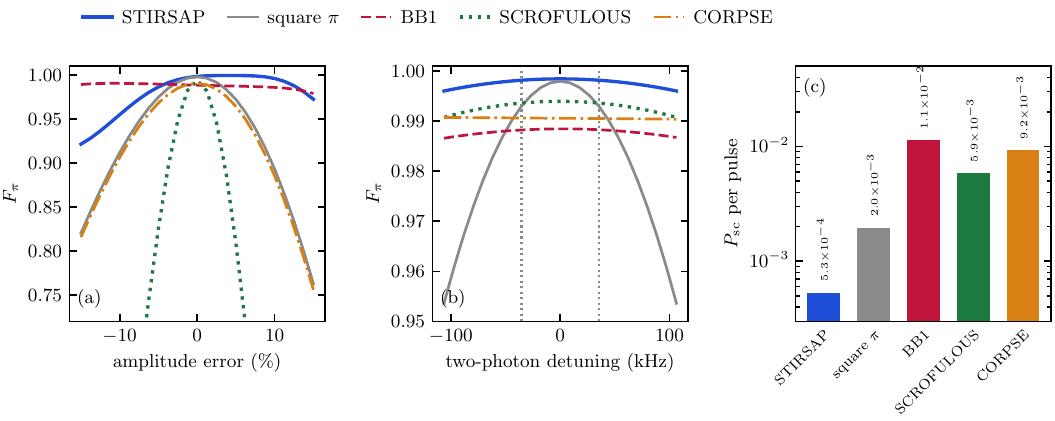}
	\caption{
		Matched-condition benchmark of STIRSAP against composite Raman pulses,
		all at $T_p=1~\mu\mathrm{s}$, $\Delta=2\pi\times5$~GHz, propagated with
		the same three-level open-system model.
		\textbf{(a)} Transfer fidelity versus amplitude (pulse-area) error; the
		SCROFULOUS curve falls below the plotted range beyond $\pm6\%$.
		\textbf{(b)} Transfer fidelity versus two-photon detuning; dotted lines
		mark the $\pm1\sigma$ Doppler width of a $2~\mu\mathrm{K}$ cloud.
		\textbf{(c)} Spontaneous-emission probability per pulse (value above each
		bar). The legend at top applies to (a) and (b); note the expanded
		vertical scale in (b). The composite sequences require three to five times
		the pulse area within the same duration and therefore scatter one to two
		orders of magnitude more.}
	\label{fig:benchmark}
\end{figure*}
\subsection{Matched-condition benchmark against composite pulses}
\label{sec:benchmark}

For the sake of completeness, we benchmark STIRSAP against
three standard composite sequences: BB1, SCROFULOUS, and CORPSE
\citep{wimperis1994broadband,cummins2003tackling}, together with a single
resonant Raman $\pi$ pulse. All protocols are propagated with the same
three-level Hamiltonian~\eqref{eq:H_lambda}, the same $\Delta=2\pi\times5$~GHz,
the same excited-state linewidth $\Gamma$, and, crucially, the same total pulse
duration $T_p=1~\mu\mathrm{s}$. Because a composite sequence of total area
$A\pi$ must fit into the same window, it runs at a correspondingly higher
instantaneous Rabi frequency; the required values are
$2\pi\times\{258,71,158,123,147\}$~MHz for
STIRSAP, square $\pi$, BB1, SCROFULOUS, and CORPSE respectively, all within the
assumed power budget.

Figure~\ref{fig:benchmark} shows the outcome. Against amplitude (pulse-area)
errors [Fig.~\ref{fig:benchmark}(a)], BB1 is the most robust, as expected from
its design, remaining above $0.987$ across $\pm15\%$; STIRSAP is second,
exceeding $0.99$ over roughly $\pm8\%$, while the single $\pi$ pulse, CORPSE,
and SCROFULOUS degrade rapidly. Against two-photon detuning
[Fig.~\ref{fig:benchmark}(b)], spanning the $\pm1\sigma$ Doppler width of a
$2~\mu\mathrm{K}$ cloud ($2\pi\times35.5$~kHz), STIRSAP, CORPSE, and SCROFULOUS
are all flat, and STIRSAP retains the highest fidelity.

The decisive difference appears in spontaneous emission
[Fig.~\ref{fig:benchmark}(c)]. Because the composite sequences must deliver
three to five times the pulse area of a single $\pi$ pulse within the same
duration, they accumulate correspondingly more excited-state population:
$P_{\mathrm{sc}}=1.1\times10^{-2}$ (BB1), $9.2\times10^{-3}$ (CORPSE), and
$5.9\times10^{-3}$ (SCROFULOUS), against $5.3\times10^{-4}$ for STIRSAP, a
factor of $11$ to $21$. In an LMT sequence with $4(n-1)$ pulses, this is the
dominant consideration: at $n=100$, BB1 would scatter essentially the entire
ensemble, whereas STIRSAP retains $\sim80\%$.

The conclusion we draw is therefore specific and limited. STIRSAP does not
achieve a higher peak single-pulse fidelity than optimized composite pulses,
and BB1 is more robust to amplitude errors. What STIRSAP provides, at matched
duration and detuning, is comparable robustness at an order of magnitude less
spontaneous emission, because it transfers population along a dark state rather
than through repeated large-area rotations. For sequential LMT interferometry,
where per-pulse scattering compounds over hundreds of pulses, this is the
relevant figure of merit.

\begin{figure*}[t]
	\centering
	\includegraphics[width=0.8\textwidth]{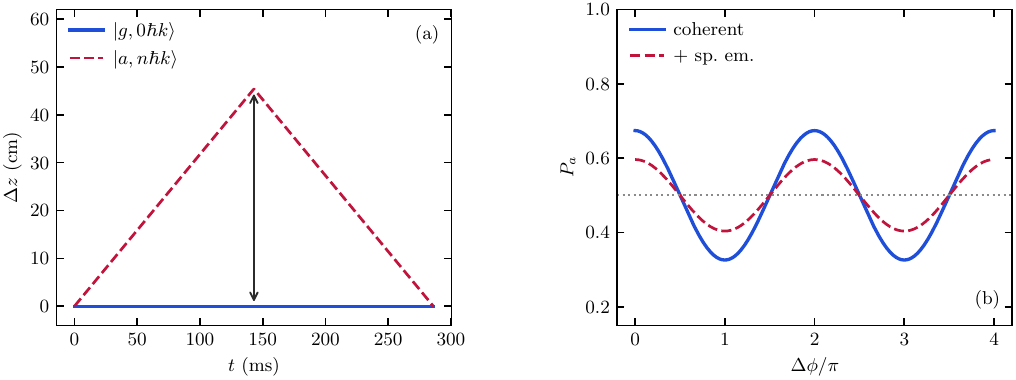}
	\caption{ LMT Mach-Zehnder interferometer at the numerically optimal momentum-transfer order. \textbf{(a)} Classical space-time trajectories for \(n=270\). Sequential STIRSAP Raman pulses generate a maximum wave-packet separation of \(\Delta z = 45.4~\mathrm{cm}\) at \(t=T\) (vertical arrow) while preserving symmetric recombination at the final beam splitter. \textbf{(b)} Simulated interference fringe at \(n=270\) obtained by scanning the recombination phase, in the coherent model (solid, \(\mathcal{C}=0.348\)) and including spontaneous emission (dashed, \(\mathcal{C}=0.193\)); the dotted line marks \(P_a=1/2\). }
	\label{fig:fringe}
\end{figure*}

\section{LMT Mach-Zehnder Interferometry}
\label{sec:lmt}

\subsection{Sequential LMT Interferometer Sequence}
\label{sec:sequence}

We consider a Raman Mach-Zehnder interferometer employing sequential LMT pulses to enhance the effective interferometer area. An interferometer of order \(n\) consists of five stages:
 (i) An initial Raman $\pi/2$ beam splitter followed by $n-1$ acceleration pulses, generating a superposition of $\ket{g,\mathbf{p}}$ and $\ket{a,\mathbf{p}+n\hbar k_{\mathrm{eff}}\hat{z}}$.
	(ii)  Free evolution under gravity for interrogation time $T$.
	(iii)  A mirror sequence composed of deceleration pulses, a central Raman $\pi$ pulse, and reacceleration pulses.
	(iv)  A second free-evolution interval of duration $T$.
	(v)  A final recombination sequence consisting of deceleration pulses followed by a Raman $\pi/2$ pulse.
\\
The total number of additional Raman $\pi$ pulses scales as $4(n-1)$, so maintaining high transfer fidelity across the sequential pulse train is essential for high-order LMT operation. In practice, the Raman beam directions and two-photon detunings must be alternated and chirped throughout the sequence to compensate for the changing Doppler shift during momentum transfer \citep{weitz1994atomic}.
For a fountain with apogee height $z_{\max}=10~\mathrm{cm}$, the launch velocity is $v_0=\sqrt{2gz_{\max}}=1.40~\mathrm{m/s}$, corresponding to an interrogation time $T=v_0/g=143~\mathrm{ms}$.
The leading interferometric phase shift is \begin{equation} \Delta\Phi = n k_{\mathrm{eff}} g T^2 + \Delta\Phi_{\mathrm{prop}} + \Delta\Phi_{\mathrm{laser}}, \label{eq:phase} \end{equation} where the additional terms represent the propagation phase and Raman laser phase contributions \citep{storey1994feynman,antoine2003quantum}.
\subsection{Space-Time Geometry and Interference Fringes} \label{sec:trajectories} 
Figure~\ref{fig:fringe}(a) illustrates the classical space-time trajectories corresponding to the numerically optimal momentum-transfer order \( n=270. \) The sequential STIRSAP Raman pulses symmetrically accelerate and decelerate the two interferometer arms, generating a maximum momentum splitting of \( 2n\hbar k_{\mathrm{eff}} \) while preserving the standard Mach-Zehnder geometry. The resulting trajectories remain mirror symmetric about the central \(\pi\)-pulse, ensuring recombination of the two wave packets at the final beam splitter despite the large momentum separation accumulated during the sequence. The principal consequence of increasing the LMT order is the rapid growth of the spatial separation between the interferometer arms. For \( n=270, \) the maximum wave-packet separation reaches approximately \( \Delta z = 45.4~\mathrm{cm}, \) more than four times larger than the assumed \(10~\mathrm{cm}\) fountain height. This large separation reflects the substantial phase enhancement responsible for the improved shot-noise sensitivity. At the same time, it highlights the geometric constraints associated with extreme-LMT operation. Realizing such trajectories would require Raman beams with large transverse extent and highly uniform wave-front quality over the entire interferometer region. Furthermore, gravity-gradient dephasing, Coriolis effects, diffraction phases, AC Stark inhomogeneities, and optical phase curvature are all expected to become progressively more significant as the arm separation increases.

Figure~\ref{fig:fringe}(b) shows the simulated interference fringe obtained by scanning the recombination phase \( \Delta\phi. \) The sinusoidal dependence confirms coherent recombination of the two interferometer paths after the complete LMT sequence. Despite the \(4(n-1)=1076\) additional Raman \(\pi\)-pulses, the interferometer retains a finite contrast of $\mathcal{C}=0.348$ in the coherent model; including the spontaneous-emission loss of Eq.~\eqref{eq:Psc} reduces this to $\mathcal{C}=0.193$. This residual contrast is determined primarily by the compound pulse fidelity discussed in Sec.~\ref{sec:single_pulse}, demonstrating that high-order STIRSAP sequences can preserve observable interference even after hundreds of sequential momentum-transfer operations. The contrast reduction nevertheless illustrates the fundamental tradeoff governing LMT interferometry: increasing \(n\) enhances the accumulated phase while simultaneously amplifying the impact of small pulse-transfer imperfections. The optimum near \( n\approx270 \) therefore emerges from the competition between these two effects rather than from geometric constraints alone. When the per-pulse spontaneous-emission loss of Sec.~\ref{sec:threelevel} is included, the model optimum shifts downward to $n\approx170$ and the corresponding shot-noise sensitivity worsens by a factor of about $1.6$, so the value $n\approx270$ obtained from the coherent model should be regarded as an upper bound.

\subsection{Compound Fidelity and Contrast Scaling}
\label{sec:compound}
\begin{figure*}[htbp]
	\centering
	\includegraphics[width=0.9\textwidth]{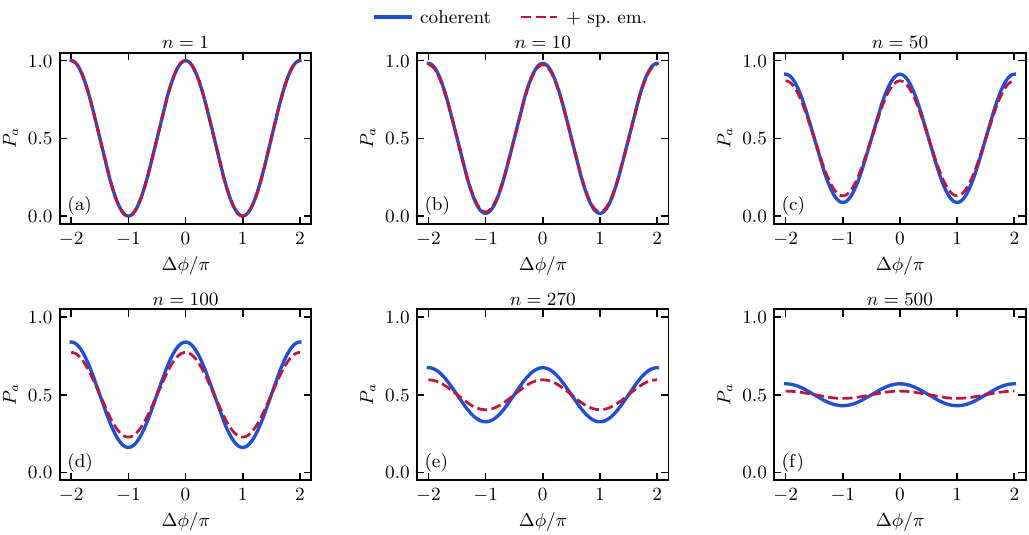}
	\caption{
		Simulated interference fringes for increasing momentum-transfer order:
		\textbf{(a)}
		\(n=1\),
		\textbf{(b)}
		\(n=10\),
		\textbf{(c)}
		\(n=50\),
		\textbf{(d)}
		\(n=100\),
		\textbf{(e)}
		\(n=270\),
		and
		\textbf{(f)}
		\(n=500\).
		The progressive reduction in fringe visibility reflects the exponential
		accumulation of pulse-transfer errors across the sequential LMT pulse train.
		Solid curves: coherent model; dashed: including spontaneous
		emission. The momentum-transfer order \(n\) is indicated above each panel,
		and the coherent/dissipative contrasts are, respectively,
		\(1.000/1.000\), \(0.965/0.946\), \(0.825/0.741\), \(0.678/0.545\),
		\(0.348/0.193\), and \(0.141/0.047\).}
	\label{fig:multi_fringes}
\end{figure*}

To estimate the cumulative effect of sequential pulse imperfections, we model the compound interferometer fidelity as $F_c(n) = F_\pi^{\,4(n-1)}$,where \(F_\pi\) is the single-pulse transfer fidelity. The corresponding interferometer contrast is approximated by 
\begin{equation} 
\mathcal{C}(n) = 2\sqrt{P_a(1-P_a)}\,F_c(n). \label{eq:contrast}
\end{equation} 
This treatment assumes that pulse imperfections act primarily as incoherent losses into unobserved momentum states. We note that for clock (single-photon) LMT interferometers, the cumulative population error from sequential inversions applied from alternating directions has been shown to scale linearly with $n$ rather than quadratically, and parasitic-path contributions to be negligible \citep{jiang2026cumulative}; the Raman geometry considered here differs, but the same distinction between compounding and merely additive errors underlies our compound-fidelity estimate. While useful as a first-order estimate, this approximation becomes increasingly optimistic at high momentum order. Coherent population leakage between diffraction channels can generate parasitic interferometer paths and additional phase shifts \citep{schkolnik2015effect}, effects not captured by the simplified exponential-fidelity model. Figures~\ref{fig:multi_fringes}(a)-(f) show the resulting fringe evolution from \(n=1\) to \(n=500\). The contrast decreases monotonically from \( \mathcal{C}=1 \) at \(n=1\) (see Fig. \ref{fig:multi_fringes}(a)) to \( \mathcal{C}\approx0.35 \) at \(n=270\) (Fig. \ref{fig:multi_fringes}(e)), and finally \( \mathcal{C}\approx0.14 \) at \(n=500\) (Fig. \ref{fig:multi_fringes}(f)), demonstrating the rapid accumulation of pulse-transfer errors in sequential LMT operation.

\subsection{Shot-Noise-Limited Sensitivity}
\label{sec:shotnoise}

Within the simplified interferometer model, the shot-noise-limited acceleration sensitivity is given by
\begin{equation} 
	\delta g = \frac{1}{n k_{\mathrm{eff}} T^2} \sqrt{ \frac{1}{\mathcal{C}^2 \eta N} + \sigma_\varphi^2 } \sqrt{T_{\mathrm{cyc}}}, \label{eq:sensitivity}
\end{equation} 
where \(N\) is the detected atom number, \(\eta\) the detection efficiency, \(\sigma_\varphi\) the residual Raman phase noise, and \(T_{\mathrm{cyc}} = 2T + (4n-1)T_p\) the total interferometer cycle time.

The achievable sensitivity is governed by the competition among three mechanisms: (i) linear phase enhancement with increasing momentum-transfer order \(n\), (ii) exponential contrast reduction arising from compound pulse infidelity, and (iii) growth of the pulse-sequence overhead at large LMT order. Because the STIRSAP Raman pulses considered here are ultrashort (\(T_p = 1~\mu\mathrm{s}\)), the pulse-overhead contribution remains negligible even at very large momentum order. For example, at \(n = 1000\) the additional Raman pulse time contributes only \(4~\mathrm{ms}\) to the cycle duration, corresponding to approximately \(1.4\%\) of the total interferometer sequence. This suppression of dead-time overhead is one of the principal advantages of shortcut-to-adiabatic Raman control relative to conventional adiabatic Raman protocols, which employ substantially longer pulse durations.

\begin{figure*}[t]
	\centering
	\includegraphics[width=0.8\textwidth]{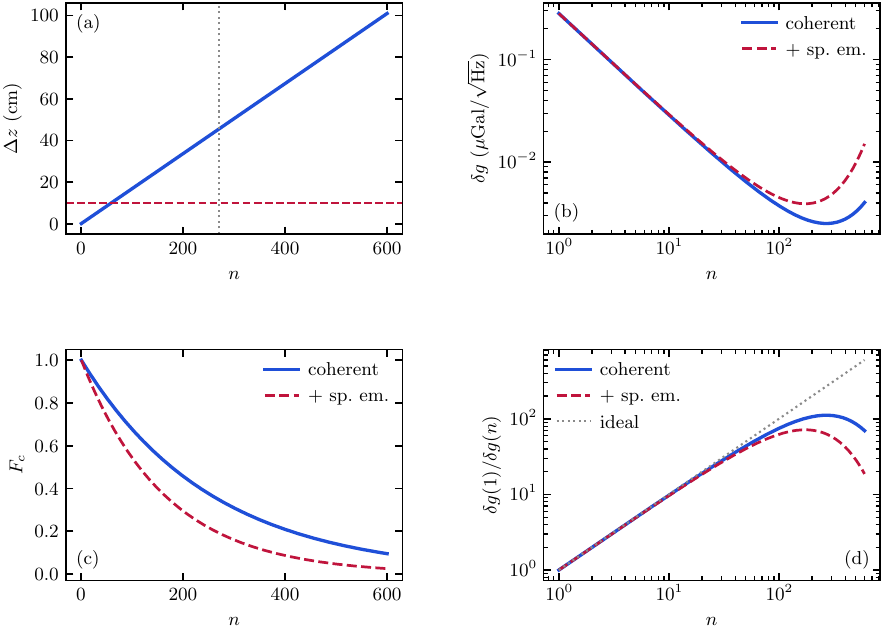}
	\caption{
		Scaling behavior of the STIRSAP-LMT interferometer with momentum-transfer
		order.
		\textbf{(a)}
		Maximum wave-packet separation versus
		\(n\),
		reaching approximately
		\(45.4~\mathrm{cm}\)
		at the unconstrained optimum
		\(n\approx270\) (dotted vertical line); the dashed horizontal line marks the
		\(10~\mathrm{cm}\) fountain height.
		\textbf{(b)}
		Shot-noise-limited sensitivity predicted by the simplified interferometer
		model. The optimum excludes vibration and other technical noise sources.
		\textbf{(c)}
		Compound fidelity
		\(F_c\)
		versus momentum-transfer order, in the coherent model and including
		spontaneous emission. The interferometer contrast follows the same
		trend up to the near-unity prefactor \(2\sqrt{P_a(1-P_a)}\) of
		Eq.~\eqref{eq:contrast}.
		\textbf{(d)}
		Sensitivity enhancement relative to the
		\(n=1\)
		configuration, compared with ideal lossless linear scaling.}
	\label{fig:scaling}
\end{figure*}

Neglecting the weak dependence of \(T_{\mathrm{cyc}}\) on \(n\), the optimal momentum-transfer order follows approximately from \(\partial(\delta g)/\partial n = 0\), yielding 
\begin{equation} n_{\mathrm{opt}} \approx \frac{1}{4\,|\ln F_\pi|} \approx 255. \label{eq:nopt} \end{equation}
This estimate agrees reasonably well with the full numerical optimization, which gives \(n_{\mathrm{opt}} = 270\) for the parameters considered here (see Fig.~\ref{fig:scaling}). The remaining discrepancy arises primarily from the weak cycle-time dependence neglected in deriving Eq.~\eqref{eq:nopt}. For \(N = 10^6\), \(\eta = 0.5\), and \(\sigma_\varphi = 10^{-3}~\mathrm{rad}\), the corresponding shot-noise floor is \(\delta g_{\mathrm{opt}} = 0.0025~\mu\mathrm{Gal}/\sqrt{\mathrm{Hz}}\). Relative to the \(n = 1\) configuration, this corresponds to a nominal \(112\times\) improvement within the simplified shot-noise model. Including the per-pulse spontaneous-emission loss of Sec.~\ref{sec:threelevel} reduces this to \(72\times\), attained at \(n\approx170\). However, this enhancement should not be interpreted as experimentally achievable within the assumed compact fountain geometry. The same optimization simultaneously predicts wave-packet separations substantially exceeding the physical fountain size, while technical noise and systematic effects dominate well before the nominal quantum-noise floor is reached. The predicted sensitivity further depends strongly on interrogation time, atom number, and pulse fidelity. Since \(\delta g \propto T^{-2}\), increasing the fountain height can substantially improve the nominal shot-noise-limited sensitivity. Similarly, increasing the detected atom number improves the sensitivity through standard projection-noise scaling, \(\delta g \propto N^{-1/2}\). Figure~\ref{fig:scaling} shows that the STIRSAP-LMT scaling initially follows the ideal linear enhancement expected for lossless LMT interferometry before compound pulse-transfer errors progressively bend the scaling curve at large momentum order.

The complete scaling behavior is summarized in Fig.~\ref{fig:scaling}. Panel~\ref{fig:scaling}(a) shows the maximum spatial (wave-packet) separation of the interferometer arms as a function of the momentum-transfer order \(n\). Owing to the linear increase of the recoil velocity with \(n\), the arm separation grows approximately linearly and reaches \(\sim 45.4~\mathrm{cm}\) at the unconstrained optimum \(n \approx 270\). Although such large separations are advantageous for inertial sensitivity, they already exceed the dimensions of most laboratory-scale fountain geometries, indicating that practical implementations will generally require additional constraints on the accessible momentum-transfer order.

The sensitivity predicted by Eq.~\eqref{eq:sensitivity} is shown in Fig.~\ref{fig:scaling}(b). For small \(n\), the dominant effect is the linear increase of the interferometric phase shift, leading to a rapid improvement in sensitivity approximately proportional to \(1/n\). As \(n\) increases further, however, the cumulative effect of imperfect Raman transfers progressively reduces the interferometer contrast, eventually offsetting the gain obtained from the larger enclosed space-time area. The resulting competition produces a broad minimum at \(n_{\mathrm{opt}} \approx 270\), corresponding to a nominal shot-noise floor of \(\delta g_{\mathrm{opt}} \approx 2.5 \times 10^{-3}~\mu\mathrm{Gal}/\sqrt{\mathrm{Hz}}\).
Figure~\ref{fig:scaling}(c) illustrates the origin of this optimum through the dependence of the compound fidelity \(F_c\) and interferometer contrast \(\mathcal{C}\) on the momentum-transfer order. For moderate \(n\), both quantities remain close to unity because of the high single-pulse STIRSAP fidelity. Beyond \(n \sim 10^2\), however, the accumulation of transfer errors over the \(4(n-1)\) Raman pulses causes an increasingly rapid degradation. The contrast closely follows the compound fidelity, confirming that pulse-transfer imperfections constitute the dominant limitation in the present model.

Finally, panel~\ref{fig:scaling}(d) of Fig.~\ref{fig:scaling} compares the sensitivity enhancement obtained with STIRSAP-LMT to the ideal lossless scaling expected from a perfectly efficient interferometer. For small and intermediate momentum orders, the STIRSAP-based sequence closely follows the ideal linear behavior, demonstrating that the high-fidelity shortcut pulses preserve most of the expected LMT gain. At larger \(n\), the enhancement begins to saturate and eventually decreases as contrast losses outweigh the benefits of additional momentum transfer. The maximum enhancement occurs near \(n_{\mathrm{opt}}\), where the STIRSAP-LMT interferometer achieves approximately two orders of magnitude improvement over the \(n = 1\) configuration while remaining close to the ideal scaling limit.

\subsection{Technical Noise and Allan Deviation}
\label{sec:allan}

\begin{figure*}[t]
	\centering
	\includegraphics[width=0.8\textwidth]{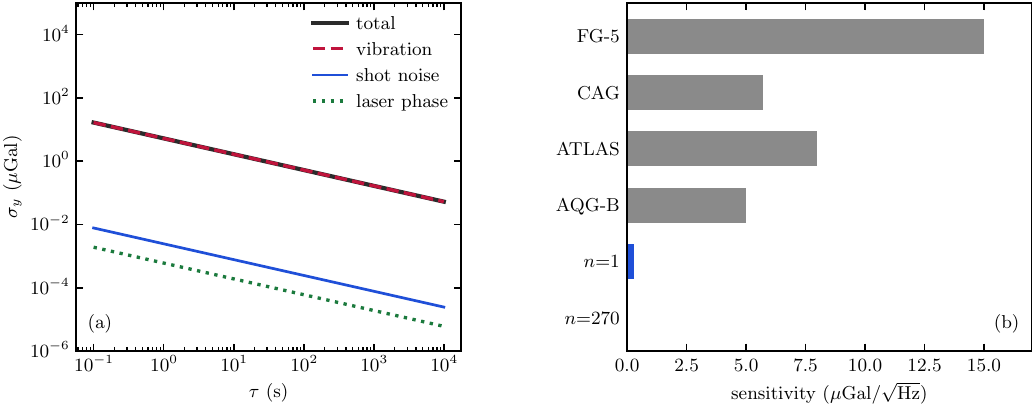}
	\caption{
		Noise and stability analysis for the STIRSAP-LMT interferometer.
		\textbf{(a)}
		Projected Allan deviation including shot noise, Raman phase noise,
		and vibration noise, each plotted separately. The vibration contribution
		(red dashed) lies almost exactly under the total (black), since it exceeds
		the other terms by more than three orders of magnitude; all curves are
		shown with distinct styles for clarity.
		\textbf{(b)}
		Comparison between the theoretical shot-noise floor predicted for the
		STIRSAP-LMT model (\(0.282\) and \(0.003~\mu\mathrm{Gal}/\sqrt{\mathrm{Hz}}\)
		at \(n=1\) and \(n=270\), respectively) and representative experimental gravimeter
		sensitivities reported in the literature (FG-5, \(15\); CAG, \(5.7\);
		ATLAS, \(8\); AQG-B, \(5~\mu\mathrm{Gal}/\sqrt{\mathrm{Hz}}\)).
		The STIRSAP value excludes dominant technical noise contributions and
		therefore should not be interpreted as a directly achievable experimental
		sensitivity; it is shown only to place the intrinsic shot-noise floor on
		the same axis as fully characterized devices.}
	\label{fig:allan}
\end{figure*}

The value \(\delta g=0.0025~\mu\mathrm{Gal}/\sqrt{\mathrm{Hz}}\) represents a theoretical shot-noise floor rather than a realistic experimental sensitivity. Under practical operating conditions, vibration noise is expected to dominate the interferometer phase stability. Assuming a residual vibration acceleration of order \( a_{\mathrm{vib}}\sim10^{-8}g, \) the corresponding acceleration noise is \(\simeq10~\mu\mathrm{Gal}\), about three and a half orders of magnitude above the nominal quantum-noise floor at the interrogation time considered here. Including vibration noise in the optimization fundamentally alters the predicted high-order scaling and shifts the practical optimum toward substantially lower momentum-transfer order, likely in the range \(n\sim10\)-\(30\). Additional systematic effects not included explicitly in the present simulations include (spontaneous emission is now treated explicitly in Sec.~\ref{sec:threelevel}) differential Doppler detuning across the sequential Raman pulse train, wave-front distortions, gravity-gradient-induced phase shifts, Coriolis effects, diffraction phases, and recoil-induced heating over the large number of Raman momentum kicks.
The projected Allan deviation is modeled as
\begin{equation}
	\sigma_y(\tau)
	=
	\sqrt{
		\sigma_{\mathrm{shot}}^2(\tau)
		+
		\sigma_{\mathrm{vib}}^2(\tau)
		+
		\sigma_{\mathrm{laser}}^2(\tau)
	},
	\label{eq:allan}
\end{equation}
where all three contributions are treated as uncorrelated shot-to-shot
processes averaging as \(\tau^{-1/2}\). We note that this is an idealization for
the vibration term: real platform vibration has substantial low-frequency
content that does not average down in this way, so the long-\(\tau\) behaviour
of the vibration curve in Fig.~\ref{fig:allan}(a) should be read as a
best case rather than a prediction.

Figure~\ref{fig:allan}(a) shows the resulting Allan-deviation estimate for
\(n=270\).
The vibration contribution dominates at short averaging times, whereas the
shot-noise contribution remains substantially below the technical-noise floor.
The long-term averaging behavior therefore depends critically on the degree to
which vibration and correlated technical noise can be suppressed in a realistic
experimental implementation.
Figure~\ref{fig:allan}(b) compares the theoretical shot-noise floor predicted
here with representative sensitivities reported for existing atom gravimeters.
This comparison should be interpreted cautiously because the published values
include full experimental noise contributions, whereas the STIRSAP-LMT result
shown here excludes dominant technical and systematic effects. A quantitatively
fair comparison would require evaluating the STIRSAP interferometer under the
same experimental noise conditions as the reported devices.

\section{Experimental Feasibility Considerations}
\label{sec:feasibility}

The present analysis is based on an effective Raman model and is intended primarily to investigate the scaling behavior of shortcut-to-adiabatic LMT atom optics. To assess the practical implications of the proposed STIRSAP-LMT protocol, it is useful to estimate the associated experimental requirements and implementation constraints. For the parameters used throughout this work, the peak Raman coupling is \( \Omega_0 = 2\pi\times250~\mathrm{MHz} \) with single-photon detuning \( \Delta = 2\pi\times5~\mathrm{GHz}. \) Assuming typical D$_2$-line dipole matrix elements for \({}^{87}\mathrm{Rb}\), these values correspond to Raman intensities of several \(\mathrm{W/cm^2}\), depending on polarization configuration and optical efficiency. At large momentum-transfer order, the dominant geometric challenge is the wave-packet separation generated during the interferometer sequence. For the unconstrained optimum \( n=270, \) the arm separation reaches approximately \(45~\mathrm{cm}\), substantially exceeding the assumed fountain geometry. Addressing both interferometer arms simultaneously would therefore require Raman beams with large spatial extent together with high wave-front uniformity across the full interaction region. Wave-front curvature and spatial intensity inhomogeneity can then introduce diffraction phases, AC Stark shifts, and contrast loss that grow rapidly with increasing momentum order \citep{zhou2016effect}. 

Sequential momentum transfer also produces progressively increasing Doppler shifts between the interferometer arms. For Raman momentum kicks of \(2\hbar k_{\mathrm{eff}}\), the velocity separation scales as \( \Delta v = 2n v_{\mathrm{rec}}, \) where \( v_{\mathrm{rec}}=5.88~\mathrm{mm/s} \) is the single-photon recoil velocity for \({}^{87}\mathrm{Rb}\). At \( n=270, \) the differential velocity exceeds \(3~\mathrm{m/s}, \) corresponding to Raman Doppler shifts on the MHz scale. Maintaining resonant Raman transfer across the pulse sequence would therefore require pulse-by-pulse frequency chirping together with precise phase synchronization. Quantitatively, each \(2\hbar k_{\mathrm{eff}}\) kick changes the two-photon Doppler shift by \(2k_{\mathrm{eff}}v_{\mathrm{rec}}=2\pi\times30~\mathrm{kHz}\), so the resonance must be tracked over a range of order \(2\pi\times8~\mathrm{MHz}\) across the full \(n=270\) sequence. Since the Fourier bandwidth of a \(1~\mu\mathrm{s}\) pulse is \(\sim2\pi\times1~\mathrm{MHz}\), the per-pulse shift is small compared with the pulse bandwidth, so a pulse-by-pulse frequency step suffices and the residual two-photon detuning error per pulse remains within the high-fidelity plateau of Fig.~\ref{fig:construction}(h). The short STIRSAP pulse duration \( (T_p=1~\mu\mathrm{s}) \) implies Fourier-limited bandwidths near the MHz scale, partially mitigating the Doppler sensitivity. However, quantitatively accurate treatment of the chirped Raman dynamics would require full momentum-space simulations beyond the scope of the present work. Although the Raman transitions operate in the far-detuned regime, spontaneous scattering accumulates over the large number of sequential Raman pulses. The full three-level simulation of Sec.~\ref{sec:threelevel} gives a scattering probability of \(5.3\times10^{-4}\) per pulse [Eq.~\eqref{eq:Psc}], about five times the value \(10^{-4}\) assumed in earlier estimates. The cumulative scattering probability, \(1-(1-P_{\mathrm{sc}})^{4(n-1)}\approx44\%\) for the full \(n=270\) interferometer sequence, then reinforces the conclusion that extreme-LMT operation is impractical in this geometry. Such scattering would reduce atom number and fringe contrast while introducing decoherence and systematic phase shifts. 

The sensitivity analysis further shows that vibration noise dominates well before the nominal shot-noise floor is reached. Residual platform acceleration at the level of \(10^{-8}g\) already exceeds the predicted quantum-noise floor by several orders of magnitude for the interrogation times considered here. Consequently, practical operation at high momentum-transfer order would require substantial vibration isolation or post-correction techniques similar to those employed in state-of-the-art atom gravimeters. Taken together, these considerations indicate that the principal limitations of extreme-LMT STIRSAP interferometry arise not from pulse duration itself, but from large spatial separation, technical noise, and systematic phase control. The results therefore suggest that shortcut-to-adiabatic Raman shaping may be most advantageous in intermediate-LMT regimes, where ultrafast pulse operation can suppress dead-time overhead while maintaining experimentally accessible interferometer geometries and technical-noise requirements.

\section{Discussion}
\label{sec:discussion}

The present work investigates shortcut-to-adiabatic Raman control as a possible route toward scalable LMT atom optics for compact interferometry. The results should therefore be interpreted within the broader context of existing high-order atom-optics approaches, including Bragg diffraction, Bloch oscillations, composite Raman pulses, and optimal-control protocols. A central feature of the STIRSAP approach is that the CD correction is implemented entirely through optical shaping of the Raman pulse envelopes, without auxiliary microwave or radio-frequency couplings \citep{du2016experimental}. The principal advantage suggested by the present simulations is therefore not necessarily higher ultimate single-pulse fidelity, but rather the combination of high-fidelity transfer with ultrashort pulse duration. Conventional adiabatic Raman protocols typically require pulse durations of order \(10\)-\(100~\mu\mathrm{s}\), which can introduce substantial dead-time overhead when extended to large momentum-transfer order. By contrast, the \(1~\mu\mathrm{s}\) STIRSAP pulses considered here contribute negligibly to the total interferometer cycle time even for long pulse sequences. At the same time, the present results do not establish that STIRSAP outperforms state-of-the-art optimized Raman-control techniques. 

Composite pulses and optimal-control methods have demonstrated transfer fidelities approaching or exceeding \(99.99\%\) under experimentally optimized conditions \citep{berg2015composite,butts2013efficient,saywell2019optimized}. The single-pulse fidelity obtained here, \( F_\pi=0.99902, \) while high, is therefore not itself beyond the current state of the art. The matched-condition benchmark of Sec.~\ref{sec:benchmark} makes the comparison explicit: at equal duration, detuning, and power budget, BB1 is more robust than STIRSAP against pulse-area errors, and several composite sequences match it against Doppler detuning. The distinguishing feature is spontaneous emission. Because composite sequences deliver three to five times the pulse area within the same window, they scatter one to two orders of magnitude more per pulse (\(1.1\times10^{-2}\) for BB1 against \(5.3\times10^{-4}\) for STIRSAP). For a single gate this is immaterial; for a sequential LMT interferometer with \(4(n-1)\) pulses it is decisive, and it is the specific sense in which superadiabatic Raman shaping is advantageous here. A comparison against numerically optimized optimal-control pulses, whose area is not fixed a priori, remains outside the present scope. Several limitations of the present model should also be emphasized. The compound-fidelity treatment assumes that pulse imperfections act primarily as incoherent loss into unobserved momentum states. In realistic high-order interferometers, coherent population leakage between diffraction channels can generate parasitic interferometer paths with distinct phases and contrasts \citep{schkolnik2015effect}. Accurate modeling of such effects will require multimode momentum-space simulations including diffraction phases, Doppler detuning, and spatial beam inhomogeneity. On the detection side, the momentum-space fringe period scales as $1/n$, so finite detector resolution imposes a separate constraint on the usable momentum-transfer order; this trade-off between large-$n$ phase enhancement and detection blur is analyzed in a solvable model in Ref.~\citep{ali2026detection}. 

The construction and the LMT sequence model are formulated in the effective two-level Raman description obtained through adiabatic elimination of the excited state. This description has been validated directly in Sec.~\ref{sec:threelevel}: full three-level simulations, including transient excited-state population and spontaneous emission, confirm the effective model to better than \(10^{-4}\) in transfer fidelity and establish the per-pulse scattering probability \(P_{\mathrm{sc}}\simeq5.3\times10^{-4}\) that sets the practical LMT optimum. More broadly, the analysis highlights an important transition in the limiting physics of high-order atom interferometry. Once Raman pulse durations become sufficiently short, the dominant constraints no longer arise from pulse overhead itself, but instead from systematic phase accumulation, technical noise, Doppler control, and large spatial separation between interferometer arms. In this sense, the work should be viewed less as a proposal for an immediately realizable extreme-LMT gravimeter and more as a quantitative study of the fidelity-limited scaling behavior of superadiabatic Raman atom optics. What the present results support is narrower than a general recommendation: at the pulse durations studied, superadiabatic Raman shaping combines sub-microsecond operation with a per-pulse scattering probability roughly an order of magnitude below that of composite sequences of equal duration. Whether that translates into an experimental advantage depends on the effects not modelled here.

\section{Conclusions} 
\label{sec:conclusions}
We have investigated theoretically the application of shortcut-to-adiabatic Raman atom optics to large-momentum-transfer (LMT) Mach-Zehnder atom interferometry in a compact fountain geometry. Using an all-optical stimulated Raman shortcut-to-adiabatic passage (STIRSAP) protocol, the CD correction is incorporated directly into the Raman pulse envelopes without auxiliary microwave or radio-frequency fields. Within an effective Raman description, the resulting \(1~\mu\mathrm{s}\) pulses achieve single-pulse transfer fidelities of \( F_\pi=0.99902, \) while maintaining negligible pulse-sequence overhead even at large momentum-transfer order. The analysis shows that ultrafast superadiabatic Raman control can substantially reduce one of the traditional limitations of adiabatic LMT atom optics, namely the accumulation of dead time associated with long pulse sequences. In the present model, the pulse overhead remains below the percent level even for \(n\sim10^3\), such that the interferometer scaling becomes governed primarily by the competition between phase enhancement and compound fidelity decay. 

Within the simplified shot-noise model considered here, the unconstrained sensitivity optimum occurs near \( n_{\mathrm{opt}}\approx270, \) corresponding to a nominal acceleration sensitivity of \( 0.0025~\mu\mathrm{Gal}/\sqrt{\mathrm{Hz}}. \) Including the spontaneous-emission loss obtained from the three-level treatment moves this optimum to \(n\approx170\) and degrades the nominal floor by a factor of about \(1.6\), so the coherent value should be read as an upper bound. At the same time, the analysis reveals that the dominant limitations of high-order interferometry in this regime arise not from pulse duration itself, but from large spatial separation, technical noise, and systematic phase accumulation. In particular, the predicted wave-packet separation at the nominal optimum substantially exceeds the compact fountain geometry considered here, while realistic vibration noise is expected to dominate the phase stability well before the nominal shot-noise floor is reached. The present work should therefore be interpreted primarily as a theoretical study of fidelity-limited scaling in superadiabatic Raman interferometry rather than as a prediction of experimentally achievable gravimeter performance. More generally, the results clarify the physical tradeoffs that emerge when shortcut-to-adiabatic Raman control is extended to extreme LMT order, including the interplay between pulse fidelity, interferometer geometry, Doppler control, and technical noise. 

Full three-level simulations and a matched-condition benchmark against composite Raman sequences have been carried out here (Secs.~\ref{sec:threelevel} and~\ref{sec:benchmark}). Future work should therefore focus on multimode momentum-space dynamics including coherent inter-channel leakage, realistic noise modeling, and benchmarking against numerically optimized optimal-control pulses, whose pulse area is not fixed a priori. Experimental implementation will additionally require advances in wave-front engineering, vibration rejection, Doppler compensation, and large-aperture Raman beam delivery before the ultimate limits of high-order STIRSAP interferometry can be quantitatively assessed.

\section*{Acknowledgements}
Research reported in this paper was supported by the Qatar Research Development
and Innovation Council [ARG01-0603-230468]. The authors thank the Qatar Center
for Quantum Computing for providing computational resources.
H.A.Z. acknowledges funding by the Slovak Research and Development Agency under the contract No. APVV-24-0091, and the financial support provided under the postdoctoral fellowship program of P. J. \v{S}af\'{a}rik University in Ko\v{s}ice, Slovakia.

\appendix

\section{Simulation Parameters}
\label{app:params}
We summarize the principal physical, Raman-control, and numerical parameters used throughout the STIRSAP-LMT simulations. The listed quantities correspond to the representative operating conditions adopted in the main text for the single-pulse analysis, LMT interferometry, and sensitivity calculations. Unless otherwise stated, all numerical results and figures were generated using the parameter values summarized in Tables~\ref{tab:feasibility} to~\ref{tab:results}.
\begin{table}[t]
	\centering
	\caption{Representative experimental scales for the STIRSAP-LMT interferometer.}
	\begin{tabular}{lc}
		\toprule
		Quantity & Estimated scale \\
		\midrule
		Peak Raman coupling & $2\pi\times250~\mathrm{MHz}$ \\
		Single-photon detuning & $2\pi\times5~\mathrm{GHz}$ \\
		Pulse duration & $1~\mu\mathrm{s}$ \\
		Maximum arm separation ($n=270$) & $45~\mathrm{cm}$ \\
		Maximum differential velocity & $\sim3~\mathrm{m/s}$ \\
		Estimated Raman Doppler shift & MHz scale \\
		Estimated scattering probability/pulse & $5.3\times10^{-4}$ \\
		Representative residual vibration level & $\sim10^{-8}g$ \\
		\bottomrule
	\end{tabular}
	\label{tab:feasibility}
\end{table}
\begin{table}[t]
	\centering
	\caption{
		Representative simulation and experimental parameters used in the
		STIRSAP-LMT analysis.}
	\label{tab:params}
	\begin{tabular}{lc}
		\toprule
		Parameter & Value \\
		\midrule
		Atomic species & ${}^{87}$Rb \\
		Peak Raman coupling & $2\pi\times250~\mathrm{MHz}$ \\
		Single-photon detuning & $2\pi\times5~\mathrm{GHz}$ \\
		STIRSAP pulse duration & $1~\mu\mathrm{s}$ \\
		Interrogation time & $143~\mathrm{ms}$ \\
		Detected atom number & $10^6$ \\
		Cloud temperature\footnote{The cloud temperature does not enter Eq.~\eqref{eq:sensitivity}, which is independent of velocity spread; it is used only to set the Doppler width in the robustness analysis of Sec.~\ref{sec:benchmark}.} & $2~\mu\mathrm{K}$ \\
		Detection efficiency & $0.5$ \\
		Residual phase noise & $10^{-3}~\mathrm{rad}$ \\
		Representative vibration level & $10^{-8}g$ \\
		\bottomrule
	\end{tabular}
\end{table}
\begin{table}[b]
	\caption{
		Principal performance metrics predicted by the STIRSAP-LMT model, where spontaneous emission changes a value; both the coherent
		and dissipative figures are given.}
	\label{tab:results}
	\begin{ruledtabular}
	\begin{tabular}{lc}
		Metric & Value \\
		\colrule
		Single-pulse fidelity $F_\pi$ & $0.99902$ \\
		Optimal LMT order $n_{\mathrm{opt}}$ & $270$ \\
		Maximum arm separation & $45~\mathrm{cm}$ \\
		Compound fidelity $F_c$ ($n{=}270$)\footnote{Coherent value; including spontaneous emission gives $0.193$.} & $0.348$ \\
		Interferometer contrast $\mathcal{C}$ & $0.348$ \\
		Cycle time $T_{\mathrm{cyc}}$ & $287.1~\mathrm{ms}$ \\
		Pulse overhead & $1.08~\mathrm{ms}$ \\
		Shot-noise floor $\delta g$ & $0.0025~\mu\mathrm{Gal}/\sqrt{\mathrm{Hz}}$ \\
		Sensitivity enhancement\footnote{Coherent value at $n{=}270$; including spontaneous emission gives $72\times$ at $n\approx170$.} & $112\times$ \\
	\end{tabular}
	\end{ruledtabular}
\end{table}

%\vspace{10cm}
\textbf{Author contributions:}
A.A. conceived the main idea, developed the theoretical framework, performed the analytical calculations and numerical simulations, interpreted the results, and prepared the first complete draft of the manuscript. All other authors contributed to the critical review and proofreading of the manuscript, including checking the mathematical derivations, physical arguments, consistency of the theoretical analysis, and verification of the reported findings, and provided feedback and revisions to improve the manuscript. All authors approved the final version of the manuscript.
\\
\\
\textbf{Data~availability:}
All data that support the findings of this study are included within the article.\\
\\
\textbf{Competing~interests:}
The authors declare that they have no competing interests.
%%%%%%%%%%%%%%%%%%%%%%%%%%%%%%%%%%%%%%%%%%%%%%%%%%%%%%%%%%%%%%%%%%%%%%%%%%%%%%%%%%%%%%%%%%%%%%%%%%%%%

%\bibliographystyle{unsrtnat}
\bibliography{references_fdp_revised}

@article{demirplak2003adiabatic,
  author    = {Demirplak, M. and Rice, S. A.},
  title     = {Adiabatic population transfer with control fields},
  journal   = {J. Phys. Chem. A},
  volume    = {107},
  pages     = {9937--9945},
  year      = {2003},
  doi       = {10.1021/jp030708a}
}

@article{berry2009transitionless,
  author    = {Berry, M. V.},
  title     = {Transitionless quantum driving},
  journal   = {J. Phys. A: Math. Theor.},
  volume    = {42},
  pages     = {365303},
  year      = {2009},
  doi       = {10.1088/1751-8113/42/36/365303}
}

@article{kasevich1991atomic,
  author    = {Kasevich, M. and Chu, S.},
  title     = {Atomic interferometry using stimulated {Raman} transitions},
  journal   = {Phys. Rev. Lett.},
  volume    = {67},
  pages     = {181--184},
  year      = {1991},
  doi       = {10.1103/PhysRevLett.67.181}
}

@article{peters1999measurement,
  author    = {Peters, A. and Chung, K. Y. and Chu, S.},
  title     = {Measurement of gravitational acceleration by dropping atoms},
  journal   = {Nature},
  volume    = {400},
  pages     = {849--852},
  year      = {1999},
  doi       = {10.1038/23655}
}

@article{peters2001high,
  author    = {Peters, A. and Chung, K. Y. and Chu, S.},
  title     = {High-precision gravity measurements using atom interferometry},
  journal   = {Metrologia},
  volume    = {38},
  pages     = {25--61},
  year      = {2001},
  doi       = {10.1088/0026-1394/38/1/4}
}

@article{menoret2018gravity,
  author    = {M\'{e}noret, V. and Vermeulen, P. and Le Moigne, N. and Bonvalot, S.
               and Bouyer, P. and Landragin, A. and Desruelle, B.},
  title     = {Gravity measurements below {$10^{-9}\,g$} with a transportable absolute quantum gravimeter},
  journal   = {Sci. Rep.},
  volume    = {8},
  pages     = {12300},
  year      = {2018},
  doi       = {10.1038/s41598-018-30608-1}
}

@article{hu2013demonstration,
  author    = {Hu, Z.-K. and Sun, B.-L. and Duan, X.-C. and Zhou, M.-K.
               and Chen, L.-L. and Zhan, S. and Zhang, Q.-Z. and Luo, J.},
  title     = {Demonstration of an ultrahigh-sensitivity atom-interferometry absolute gravimeter},
  journal   = {Phys. Rev. A},
  volume    = {88},
  pages     = {043610},
  year      = {2013},
  doi       = {10.1103/PhysRevA.88.043610}
}

@article{gillot2016lne,
  author    = {Gillot, P. and Francis, O. and Landragin, A. and {Pereira Dos Santos}, F.
               and Merlet, S.},
  title     = {The {LNE-SYRTE} cold atom gravimeter},
  journal   = {Metrologia},
  volume    = {51},
  pages     = {L15--L17},
  year      = {2014},
  doi       = {10.1088/0026-1394/51/5/L15}
}

@article{freier2016mobile,
  author    = {Freier, C. and Hauth, M. and Schkolnik, V. and Leykauf, B.
               and Schilling, M. and Wziontek, H. and Scherneck, H.-G.
               and M\"{u}ller, J. and Peters, A.},
  title     = {Mobile quantum gravity sensor with unprecedented stability},
  journal   = {J. Phys.: Conf. Ser.},
  volume    = {723},
  pages     = {012050},
  year      = {2016},
  doi       = {10.1088/1742-6596/723/1/012050}
}

@article{bongs2019taking,
  author    = {Bongs, K. and others},
  title     = {Taking atom interferometric quantum sensors from the laboratory to real-world applications},
  journal   = {Nat. Rev. Phys.},
  volume    = {1},
  pages     = {731--739},
  year      = {2019},
  doi       = {10.1038/s42254-019-0117-4}
}

@article{geiger2020high,
  author    = {Geiger, R. and Landragin, A. and Merlet, S. and {Pereira Dos Santos}, F.},
  title     = {High-accuracy inertial measurements with cold-atom sensors},
  journal   = {AVS Quantum Sci.},
  volume    = {2},
  pages     = {024702},
  year      = {2020},
  doi       = {10.1116/5.0009093}
}

@article{tino2021testing,
  author    = {Tino, G. M.},
  title     = {Testing gravity with cold atom interferometry: results and prospects},
  journal   = {Quantum Sci. Technol.},
  volume    = {6},
  pages     = {024014},
  year      = {2021},
  doi       = {10.1088/2058-9565/abd840}
}

@article{storey1994feynman,
  author    = {Storey, P. and Cohen-Tannoudji, C.},
  title     = {The {Feynman} path integral approach to atomic interferometry: a tutorial},
  journal   = {J. Phys. II France},
  volume    = {4},
  pages     = {1999--2027},
  year      = {1994},
  doi       = {10.1051/jp2:1994103}
}

@article{antoine2003quantum,
  author    = {Antoine, Ch. and Bord\'{e}, Ch. J.},
  title     = {Quantum theory of atomic clocks and gravito-inertial sensors: an update},
  journal   = {J. Opt. B: Quantum Semiclassical Opt.},
  volume    = {5},
  pages     = {S199--S207},
  year      = {2003},
  doi       = {10.1088/1464-4266/5/2/385}
}

@article{mcguirk2000large,
  author    = {McGuirk, J. M. and Snadden, M. J. and Kasevich, M. A.},
  title     = {Large area light-pulse atom interferometry},
  journal   = {Phys. Rev. Lett.},
  volume    = {85},
  pages     = {4498--4501},
  year      = {2000},
  doi       = {10.1103/PhysRevLett.85.4498}
}

@article{chiow2011large,
  author    = {Chiow, S.-w. and Kovachy, T. and Chien, H.-C. and Kasevich, M. A.},
  title     = {{$102\hbar k$} large area atom interferometers},
  journal   = {Phys. Rev. Lett.},
  volume    = {107},
  pages     = {130403},
  year      = {2011},
  doi       = {10.1103/PhysRevLett.107.130403}
}

@article{chiow2011102,
  author    = {Chiow, S.-w. and Kovachy, T. and Chien, H.-C. and Kasevich, M. A.},
  title     = {{$102\hbar k$} large area atom interferometers},
  journal   = {Phys. Rev. Lett.},
  volume    = {107},
  pages     = {130403},
  year      = {2011},
  doi       = {10.1103/PhysRevLett.107.130403}
}

@article{clade2009large,
  author    = {Clad\'{e}, P. and Guellati-Kh\'{e}lifa, S. and Nez, F. and Biraben, F.},
  title     = {Large momentum beam splitter using {Bloch} oscillations},
  journal   = {Phys. Rev. Lett.},
  volume    = {102},
  pages     = {240402},
  year      = {2009},
  doi       = {10.1103/PhysRevLett.102.240402}
}

@article{kotru2015large,
  author    = {Kotru, K. and Butts, D. L. and Kinast, J. M. and Stoner, R. E.},
  title     = {Large-area atom interferometry with frequency-swept {Raman} adiabatic passage},
  journal   = {Phys. Rev. Lett.},
  volume    = {115},
  pages     = {103001},
  year      = {2015},
  doi       = {10.1103/PhysRevLett.115.103001}
}

@article{rudolph2020large,
  author    = {Rudolph, J. and Wilkason, T. and Nantel, M. and Swan, H.
               and Holland, C. M. and Jiang, Y. and Garber, B. E.
               and Carman, S. P. and Hogan, J. M.},
  title     = {Large momentum transfer clock atom interferometry on the 689\,nm intercombination line of strontium},
  journal   = {Phys. Rev. Lett.},
  volume    = {124},
  pages     = {083604},
  year      = {2020},
  doi       = {10.1103/PhysRevLett.124.083604}
}

@article{berg2015composite,
  author    = {Berg, P. and Abend, S. and Tackmann, G. and Schubert, C.
               and Giese, E. and Schleich, W. P. and Narducci, F. A.
               and Ertmer, W. and Rasel, E. M.},
  title     = {Composite-light-pulse technique for high-precision atom interferometry},
  journal   = {Phys. Rev. Lett.},
  volume    = {114},
  pages     = {063002},
  year      = {2015},
  doi       = {10.1103/PhysRevLett.114.063002}
}

@article{butts2013efficient,
  author    = {Butts, D. L. and Kinast, J. M. and Timmons, B. P. and Stoner, R. E.},
  title     = {Efficient broadband {Raman} pulses for large-area atom interferometry},
  journal   = {J. Opt. Soc. Am. B},
  volume    = {30},
  pages     = {922--927},
  year      = {2013},
  doi       = {10.1364/JOSAB.30.000922}
}

@article{muller2008atom,
  author    = {M\"{u}ller, H. and Chiow, S.-w. and Long, Q. and Herrmann, S. and Chu, S.},
  title     = {Atom interferometry with up to 24-photon-momentum-transfer beam splitters},
  journal   = {Phys. Rev. Lett.},
  volume    = {100},
  pages     = {180405},
  year      = {2008},
  doi       = {10.1103/PhysRevLett.100.180405}
}

@article{cronin2007atom,
  author    = {Cronin, A. D. and Schmiedmayer, J. and Pritchard, D. E.},
  title     = {Atom interferometers},
  journal   = {Rev. Mod. Phys.},
  volume    = {81},
  pages     = {1051--1129},
  year      = {2009},
  doi       = {10.1103/RevModPhys.81.1051}
}

@article{chiow2009noise,
  author    = {Chiow, S.-w. and Herrmann, S. and Chu, S. and M\"{u}ller, H.},
  title     = {Noise-immune conjugate large-area atom interferometers},
  journal   = {Phys. Rev. Lett.},
  volume    = {103},
  pages     = {050402},
  year      = {2009},
  doi       = {10.1103/PhysRevLett.103.050402}
}

@article{bergmann1998coherent,
  author    = {Bergmann, K. and Theuer, H. and Shore, B. W.},
  title     = {Coherent population transfer among quantum states of atoms and molecules},
  journal   = {Rev. Mod. Phys.},
  volume    = {70},
  pages     = {1003--1025},
  year      = {1998},
  doi       = {10.1103/RevModPhys.70.1003}
}

@article{vitanov2017stimulated,
  author    = {Vitanov, N. V. and Rangelov, A. A. and Shore, B. W. and Bergmann, K.},
  title     = {Stimulated {Raman} adiabatic passage in physics, chemistry, and beyond},
  journal   = {Rev. Mod. Phys.},
  volume    = {89},
  pages     = {015006},
  year      = {2017},
  doi       = {10.1103/RevModPhys.89.015006}
}

@article{weitz1994atomic,
  author    = {Weitz, M. and Young, B. C. and Chu, S.},
  title     = {Atomic interferometer based on adiabatic population transfer},
  journal   = {Phys. Rev. Lett.},
  volume    = {73},
  pages     = {2563--2566},
  year      = {1994},
  doi       = {10.1103/PhysRevLett.73.2563}
}

@article{guery2019shortcuts,
  author    = {Gu\'{e}ry-Odelin, D. and Ruschhaupt, A. and Kiely, A.
               and Torrontegui, E. and Mart\'{i}nez-Garaot, S. and Muga, J. G.},
  title     = {Shortcuts to adiabaticity: concepts, methods, and applications},
  journal   = {Rev. Mod. Phys.},
  volume    = {91},
  pages     = {045001},
  year      = {2019},
  doi       = {10.1103/RevModPhys.91.045001}
}

@article{du2016experimental,
  author    = {Du, Y.-X. and Liang, Z.-T. and Li, Y.-C. and Yue, X.-X.
               and Lv, Q.-X. and Huang, W. and Chen, X. and Yan, H. and Zhu, S.-L.},
  title     = {Experimental realization of stimulated {Raman} shortcut-to-adiabatic passage with cold atoms},
  journal   = {Nat. Commun.},
  volume    = {7},
  pages     = {12479},
  year      = {2016},
  doi       = {10.1038/ncomms12479}
}

@article{du2017geometric,
  author    = {Du, Y.-X. and Yue, X.-X. and Liang, Z.-T. and Li, J.-Z.
               and Yan, H. and Zhu, S.-L.},
  title     = {Geometric atom interferometry with shortcuts to adiabaticity},
  journal   = {Phys. Rev. A},
  volume    = {95},
  pages     = {043608},
  year      = {2017},
  doi       = {10.1103/PhysRevA.95.043608}
}

@article{saywell2019optimized,
  author    = {Saywell, J. and Carey, M. and Belal, M. and Kuprov, I. and Freegarde, T.},
  title     = {Optimised {Raman} pulses for atom interferometry},
  journal   = {J. Phys. B: At. Mol. Opt. Phys.},
  volume    = {53},
  pages     = {085003},
  year      = {2020},
  doi       = {10.1088/1361-6455/ab6df6}
}

@article{schkolnik2015effect,
  author    = {Schkolnik, V. and Leykauf, B. and Hauth, M. and Freier, C. and Peters, A.},
  title     = {The effect of wavefront aberrations in atom interferometry},
  journal   = {Appl. Phys. B},
  volume    = {120},
  pages     = {311--316},
  year      = {2015},
  doi       = {10.1007/s00340-015-6138-5}
}

@article{zhou2016effect,
  author    = {Zhou, M.-K. and others},
  title     = {Effect of wavefront aberrations on the performance of cold atom gravimeters},
  journal   = {Phys. Rev. A},
  volume    = {93},
  pages     = {043610},
  year      = {2016},
  doi       = {10.1103/PhysRevA.93.043610}
}

@article{xu2023quantum,
  author  = {Xu, Hao and Song, Xue-Ke and Wang, Dong and Ye, Liu},
  title   = {Quantum sensing of control errors in three-level systems by coherent control techniques},
  journal = {Sci. China Phys. Mech. Astron.},
  volume  = {66},
  pages   = {240314},
  year    = {2023},
  doi     = {10.1007/s11433-022-2034-5}
}

@article{PhysRevA.111.012621,
  title = {Shortcuts to adiabatic transducers in a hybrid spin-superconducting system},
  author = {Zhu, Xing-Yu and Zhu, Le-Tian and Guo, Guang-Can and Tu, Tao and Li, Chuan-Feng},
  journal = {Phys. Rev. A},
  volume = {111},
  issue = {1},
  pages = {012621},
  numpages = {14},
  year = {2025},
  month = {Jan},
  publisher = {American Physical Society},
  doi = {10.1103/PhysRevA.111.012621},
  url = {https://link.aps.org/doi/10.1103/PhysRevA.111.012621}
}

@article{wimperis1994broadband,
  author  = {Wimperis, Stephen},
  title   = {Broadband, Narrowband, and Passband Composite Pulses for Use in Advanced {NMR} Experiments},
  journal = {J. Magn. Reson. Ser. A},
  volume  = {109},
  number  = {2},
  pages   = {221--231},
  year    = {1994}
}

@article{cummins2003tackling,
  author  = {Cummins, Holly K. and Llewellyn, Gavin and Jones, Jonathan A.},
  title   = {Tackling systematic errors in quantum logic gates with composite rotations},
  journal = {Phys. Rev. A},
  volume  = {67},
  number  = {4},
  pages   = {042308},
  year    = {2003},
  doi     = {10.1103/PhysRevA.67.042308}
}

@article{ali2026brightstate,
  author  = {Ali, Asad and Al-Kuwari, Saif and Hussain, M.~I. and Kuniyil, H. and Rahim, M.~T. and Haddadi, Saeed },
  title   = {Bright-state source cancellation in dissipative shortcut {Raman} atom optics},
  journal = {arXiv preprint},
  volume  = {arXiv:2606.24939},
  year    = {2026},
  eprint  = {2606.24939},
  archivePrefix = {arXiv},
  primaryClass  = {quant-ph}
}

@article{ali2026detection,
  author  = {Ali, Asad and Al-Kuwari, Saif},
  title   = {Detection-resolution limits of large-momentum-transfer atom gravimetry},
  journal = {arXiv preprint},
  volume  = {arXiv:2607.28749},
  year    = {2026},
  eprint  = {2607.28749},
  archivePrefix = {arXiv},
  primaryClass  = {quant-ph}
}

@article{shi2021effect,
  author  = {Shi, X. and Zhao, H. Q.},
  title   = {Effect of spontaneous emission on the shortcut to adiabaticity in three-state systems},
  journal = {Phys. Rev. A},
  volume  = {104},
  number  = {5},
  pages   = {052221},
  year    = {2021},
  publisher = {APS}
}

@article{jiang2026cumulative,
  author  = {Jiang, Yijun and Rudolph, Jan and Hogan, Jason M.},
  title   = {Cumulative fidelity of large-momentum-transfer clock atom interferometers in the presence of laser noise},
  journal = {Phys. Rev. A},
  volume  = {113},
  pages   = {053316},
  year    = {2026},
  doi     = {10.1103/tqrf-42yp}
}

\end{document}